\documentclass{article}

\usepackage{arxiv}

\AtBeginDocument{%
  \providecommand\BibTeX{{%
    \normalfont B\kern-0.5em{\scshape i\kern-0.25em b}\kern-0.8em\TeX}}}

\usepackage{amsmath,amsfonts}
\usepackage[font=footnotesize]{subcaption}
\usepackage{algorithmic}
\usepackage{graphicx}
\usepackage{textcomp}
\usepackage{xcolor}
\usepackage{booktabs} %
\usepackage[ruled,linesnumbered]{algorithm2e}
\usepackage{amsfonts}
\usepackage{xspace}
\usepackage{multirow}
\usepackage{lscape}
\usepackage{courier}
\usepackage{listings}
\usepackage{linegoal}
\usepackage{url}

\usepackage{soul}
\usepackage{enumitem}
\setlist{leftmargin=5mm}

\def\dfj{\textsc{Defects4J}\xspace}

\definecolor{myblue2}{RGB}{0, 43, 136}

\newcommand{\name}{Seshat\xspace}

\newcommand{\new}[1]{{\color{black}#1}}
\definecolor{myblue}{RGB}{1, 58, 99}
\definecolor{mygreen}{RGB}{158,204,183}
\definecolor{myorange}{RGB}{255,194,141}
\newcommand{\add}[1]{\textcolor{black}{#1}}

\definecolor{pblue}{rgb}{0.13,0.13,1}
\definecolor{pgreen}{rgb}{0,0.5,0}
\definecolor{pred}{rgb}{0.9,0,0}
\definecolor{pgrey}{rgb}{0.46,0.45,0.48}
\definecolor{light-gray}{gray}{0.85}

  \newsavebox{\mylisting}
  \makeatletter
  \newcommand{\lstInline}[2][,]{%
    \begingroup%
    \lstset{#1}%
    \begin{lrbox}{\mylisting}\lstinline!#2!\end{lrbox}%
    \setlength{\@tempdima}{\linegoal}%
    \ifdim\wd\mylisting>\@tempdima\hfill\\\fi%
    \lstinline!#2!%
    \endgroup%
  }
  \makeatother

\title{Predictive Mutation Analysis via Natural Language Channel in Source Code}

\author{
Jinhan Kim\\
School of Computing, KAIST\\
Republic of Korea\\
\texttt{jinhankim@kaist.ac.kr}
\And
Juyoung Jeon\\
Handong Global University\\
Republic of Korea\\
\texttt{juyoungjeon@handong.edu}
\And
Shin Hong\\
Handong Global University\\
Republic of Korea\\
\texttt{hongshin@handong.edu}
\And
Shin Yoo\\
KAIST\\
Republic of Korea\\
\texttt{shin.yoo@kaist.ac.kr}
}

\begin{document}
\maketitle

\begin{abstract}

    Mutation analysis can provide valuable insights into both System Under Test
    (SUT) and its test suite. However, it is not scalable due to the cost of
    building and testing a large number of mutants. Predictive Mutation Testing
    (PMT) has been proposed to reduce the cost of mutation testing, but it can
    only provide statistical inference about whether a mutant will be killed or
    not by the entire test suite. We propose \name, a Predictive Mutation
    Analysis (PMA) technique that can accurately predict \emph{the entire kill
    matrix}, not just the mutation score of the given test suite. \name exploits
    the natural language channel in code, and learns the relationship between
    the syntactic and semantic concepts of each test case and the mutants it can
    kill, from a given kill matrix. The learnt model can later be used to
    predict the kill matrices for subsequent versions of the program, even after
    both the source and test code have changed significantly. Empirical
    evaluation using the programs in the \dfj shows that \name can predict kill
    matrices with the average F-score of 0.83 for versions that are up to years
    apart. This is an improvement of F-score by 0.14 and 0.45 point over the
    state-of-the-art predictive mutation testing technique, and a simple
    coverage based heuristic, respectively. \name also performs as well as PMT
    for the prediction of the mutation score only. \add{When applied to a
    Mutation Based Fault Localisation (MBFL) technique, the predicted kill matrix by
    \name is successfully used to locate faults within the top ten position, showing
    its usefulness beyond prediction of mutation scores.} Once \name trains its
    model using a concrete mutation analysis, the subsequent predictions made by
    \name are on average 39 times faster than actual test-based analysis.
    \add{We also show that \name can be successfully applied to automatically
    generated test cases with an experiment using EvoSuite.}

\end{abstract}


\section{Introduction}
\label{sec:introduction}

A long standing issue in a mutation analysis is its limited
scalability~\cite{Papadakis2019aa}.
As the size of the System Under Test (SUT) grows, the 
number of generated mutants also increases significantly. Each mutant then has 
to be compiled, and executed, to check whether it is killed (i.e., detected as 
behaving differently) by any of the existing test cases, resulting in a 
significant, sometimes infeasible, amount of cost. Many different approaches 
have been proposed to improve the scalability of mutation testing, but they 
either require more complicated program instrumentation to detect internal 
state deviation instead of propagated external behaviour (weak mutation~\cite{
howden82,off_lee94}), need more sophisticated code mutation that combines 
multiple mutants into a single compilation (meta-mutation~\cite{
untch:mutation}), or simply discard some mutants (mutant sampling~\cite{
Papadakis2010sf,Gopinath2015nu}). In many of these approaches, the improvement 
in scalability is linearly bound to the number of mutants (not) analysed.

Recently, Predictive Mutation Testing (PMT) has been proposed to attack the 
scalability issue in mutation testing from a very different angle~\cite{
Zhang2018gq,Mao2019ur}. Instead of reducing the number of mutants to analyse, 
PMT collects test suite level dynamic features that are highly relevant to 
whether a mutant can be killed or not (such as the number of tests that cover
the mutated statement, or the number of times the mutated statement is executed
by the test suite), and performs statistical inference about the probability of
the mutant being killed by the given test suite.  Given sufficient preceding
mutation testing results, PMT trains a model that can predict whether a
mutant will 
be killed by a test suite. While PMT can achieve cost saving that is not 
linearly bound to the number of mutants considered, its limitation is the fact 
that it can only make a test suite level prediction, which is sufficient to 
predict the mutation score (i.e., the ratio of killed mutants to the generated 
mutants) but not the relationship between a mutant and a single test case.

This paper proposes \name\footnote{Seshat is an Egyptian deity responsible for 
writing and record keeping.}, a predictive model for the relationship between 
mutants and individual test cases. Compared to PMT, \name can predict the 
entire kill matrix\footnote{Given $m$ mutants and $n$ test cases, a kill 
matrix $M$ is an $m$-by-$n$ matrix, where $a_{ij}$ is 1 if the mutant $m_i$ is 
killed by the test $t_j$, and 0 otherwise.} that results from mutation 
analysis. We refer to this new type of predictive modelling as Predictive 
Mutation Analysis (PMA), to emphasise the finer granularity of the prediction, 
as opposed to PMT, whose outcome is the test suite level mutation analysis.

\name exploits the Natural Language (NL) channel in software~\cite{
Casalnuovo2020oj}. Natural Language channel refers to the communication 
channel that explains the conceptual contexts of the actual executions via 
natural language elements in source code, such as identifier names. In 
comparison, the traditional mutation analysis depends on the Algorithmic (AL) 
channel, via which the semantics of the program (and, naturally, its mutant) 
is actually computed. The existence of NL channel as a human to human 
communication medium has been known for a long time~\cite{Knuth1984ci}, but 
recent advances in the application of language models to source code have 
revealed that the NL channel can be a rich source of information for various 
automated tasks, such as code completion~\cite{Hindle:2012kq} or fault 
localisation~\cite{Ray2016lz}.

Consider a test case whose name contains 
domain specific terms, such as \lstinline+testAccountBalance+. We posit that 
this test case has a much higher chance of killing any mutants that are 
generated within the scope of a method named \lstinline+getAccountBalance()+ than another method named \lstinline+updateEmailAddress(int userID, String emailAddress)+. Given a kill matrix that 
has been obtained from actual mutation analysis, \name can learn the 
relationship through the similarity between the names of each test case and the mutants it can kill. Since the relationship is learnt in the NL channel, it 
can later be used to predict the relationship between unseen test cases and 
mutants, without any execution.

\new{In addition to names, we extract the syntactic and semantic concepts in
source code and test cases, using Deep Neural Network (DNN) with the word
embedding layers and the bidirectional GRUs~\cite{cho2014learning}.} We include
the change caused by the mutation, and the type of mutation, as features of our
model. Under the cross-version scenario, we evaluate \name using different
versions of subject programs in the \dfj benchmark and two mutation tools
PIT~\cite{Coles2016ft} and Major~\cite{just2014major}. \new{Although \name
solely depends on the static features, the results show that \name can predict
kill matrices with up to F-score of 0.94 and outperforms PMT and a
coverage based baseline model}. Notably, \name does more
than simply memorising the relationship between test cases and mutant locations,
as it shows an average prediction F-score of 0.78 for newly added test cases.
Compared to generating full kill matrix by executing all individual tests
against mutants, the prediction by \name is orders of magnitude faster:
encouragingly, the bigger the target program is, the higher the speed-up
becomes. The contributions of this paper are as follows:

\begin{itemize}
\item We introduce \name, a Predictive Mutation Analysis (PMA) technique that can 
predict full kill matrices for unseen mutants and test cases.

\item We formulate predictive modelling of mutation analysis as a machine 
learning problem in the Natural Language (NL) channel in source code. To our knowledge, this is the first attempt to analyse mutation results 
using the NL channel.

\item We conduct a large scale evaluation of \name using multiple versions of
real world Java projects in the \dfj benchmark, and two widely-used mutation
tools. \name achieves F-score of 0.83 on average, between versions that are
years apart. Moreover, \name outperforms an existing Predictive Mutation Testing
(PMT) technique with finer granularity of the prediction and shows comparable
results with PMT when it is used to predict the mutation scores.

\item \add{Beyond predicting the mutation score, we apply \name to Mutation Based Fault
Localisation (MBFL) technique and evaluate its localisation effectiveness on 220 buggy
programs in \dfj. It successfully locates faults within the top ten position, which
is competitive results with original MBFL technique that uses an intact kill matrix.}

\item \add{We evaluate whether \name can be applied to automatically generated
test cases. Our experimentation with EvoSuite suggests that, as long as a
meaningful naming convention is upheld during the generation of the test cases,
\name can exploit the information in the same way and predict kill matrices of
EvoSuite generated test suites with F-score of up to 0.86.}

\end{itemize}

The rest of the paper is organised as follows. Section~\ref{sec:approach}
describes how \name formulates predictive mutation analysis using the Natural
Language channel. Section~\ref{sec:experimental_setup} describes the details
about the experimental setup, and introduces the research questions.
Section~\ref{sec:result} presents and discusses the results of empirical
evaluation. Section~\ref{sec:discussion} presents discussions about changes of test quality, data
imbalance, and ablation study. Section~\ref{sec:threats} considers threats to validity, and
Section~\ref{sec:related_work} describes the related work. Finally,
Section~\ref{sec:conclusion} concludes.

\section{\name: Predictive Mutation Analysis using NL Channel}
\label{sec:approach}

This section describes how we formulate prediction of the relationship between 
mutants and test cases via Natural Language (NL) channel. It 
also presents our model architecture for prediction of kill matrices.

\subsection{Looking at Mutation through Natural Language Channel}
\label{sec:nl_channel}

The essential steps of mutation analysis are as follows. First, we mutate the
target program using syntactic transformations, i.e., mutation operators.
Second, we execute the available test cases against the mutated program.
Finally, we check whether the mutant is \emph{killed}, i.e., whether the program
behaves differently when mutated and executed by the given test cases. The
details of these steps are captured by the PIE theory~\cite{Voas:1992uq}: for a
test case to kill a mutant, it should first \textbf{E}xecute the mutant; the
execution should result in an \textbf{I}nfected internal state, which should be
\textbf{P}ropagated to the observable output.

All three stages of PIE take place in what Casalnuovo et al. call Algorithm 
(AL) channel, as opposed to Natural Language (NL) channel~\cite{
Casalnuovo2020oj}; in the source code, the AL channel represents the 
computational semantics and executions, whereas the NL channel represents the 
identifiers and comments that assist human comprehension. \new{Most of the 
existing analyses operate within the AL channel that dictates whether PIE 
conditions are satisfied or not, while the NL channel has not been considered 
as an important factor.}

An interesting recent advance, Predictive Mutation Testing (PMT), aims to build
a predictive model using the features based on the PIE
theory~\cite{Zhang2018gq,Mao2019ur}. The two dynamic features, which are known
to be the most important features in PMT, are related to the test execution:
\textit{numTestCovered} (the number of tests in the whole test suite covering
the mutated statement) and \textit{numExecuteCovered} (the number of times the
mutated statement is executed by the whole test suite). The higher these 
dynamic feature values are, the more likely it is that the PIE conditions are 
met. We note that these dynamic features are essentially statistical 
aggregation within the AL channel, and also that the static features in PMT 
all concern structural properties of the code that exist in the AL channel.

We propose to reconstruct the results of mutation analysis in the NL channel. 
Intuitively, the prediction made by PMT using the AL channel is that ``if a 
mutant is executed frequently by many different test cases, it is more likely 
to be killed''. Here, the code coverage is used as a surrogate measure of 
proximity by computational semantics, as is often the case in regression 
testing optimisation~\cite{Yoo:2010fk}. \new{Our parallel intuition is as 
follows: \emph{``if the properties of a mutant are syntactically/semantically 
similar, or closely related to, those of a given test case, it is more likely 
to be killed by the test case.''}}

\subsection{Predictive Mutation Analysis}
\label{sec:pma}

A clear benefit of using the NL channel is that we can make predictions about a
single mutant and a single test case. On the contrary, PMT depends on the
aggregation of the features over \emph{the entire test cases} and the satisfaction of
the PIE conditions. \new{We hypothesize that this will hinder making accurate predictions
for a single test case.} However, the prediction within the NL channel can be
made by learning one-to-one relationship between the mutants and the test cases.
The one-to-one relationship allows us to predictively build
the entire kill matrix, which is required by many applications of mutation
analysis such as fault localisation~\cite{Papadakis:2015sf,Moon:2014ly,
Hong:2015db, kim2021issre}, test data
generation~\cite{Kim2018gd,Papadakis2010sf,Harman2011pi}, and automated program
repair~\cite{Debroy2010oa,Debroy2014nf,Weimer2013ma,ghanbari2019practical}. To distinguish the
difference in prediction granularity, we call the one-to-one predictive
modelling of mutation results as a Predictive Mutation Analysis (PMA).

\subsection{Input Features of \name}
\label{sec:features}

\name uses the following features to perform PMA via the NL channel in the 
source code.

\subsubsection{Test and Source Method Name}

\begin{figure}[!t]
    \begin{lstlisting}[
        basicstyle=\footnotesize\ttfamily        
    ]
     public void testFactory_daysBetween_RPartial_MonthDay() {
        MonthDay start1 = new MonthDay(2, 1);
        MonthDay start2 = new MonthDay(2, 28);
        MonthDay end1 = new MonthDay(2, 28);
        MonthDay end2 = new MonthDay(2, 29);
        
        assertEquals(27, Days.daysBetween(start1, end1).getDays());
        assertEquals(28, Days.daysBetween(start1, end2).getDays());
        assertEquals(0, Days.daysBetween(start2, end1).getDays());
        assertEquals(1, Days.daysBetween(start2, end2).getDays());
        
        assertEquals(-27, Days.daysBetween(end1, start1).getDays());
        assertEquals(-28, Days.daysBetween(end2, start1).getDays());
        assertEquals(0, Days.daysBetween(end1, start2).getDays());
        assertEquals(-1, Days.daysBetween(end2, start2).getDays());    
    }
    \end{lstlisting}
    \caption{Example test method of joda-time}
    \label{fig:test_method_name_ex1}
\end{figure}

\begin{figure}[th]
    \begin{lstlisting}[
        basicstyle=\footnotesize\ttfamily
    ]
        public static double[] nullToEmpty(final double[] array) {
            if (array == null || array.length == 0) {
                return EMPTY_DOUBLE_ARRAY;
            }
            return array;
        }
    \end{lstlisting}
    \caption{Example source method of commons-lang}
    \label{fig:mutated_stmt_ex}
\end{figure}

Test cases are written to target specific parts of the source code: any 
mutants generated in the corresponding part are, by default, more likely to be 
killed by those test cases. We exploit the fact that developers often put 
meaningful names to both source and test code~\cite{ToTTestname}. By exploiting the 
linguistic links between them, we define two input features for the name of 
the test and the source method, respectively.

Figure~\ref{fig:test_method_name_ex1} shows a test method of joda-time. From 
its method name, \lstInline{testFactory_daysBetween_RPartial_MonthDay}, we can
deduce that it is likely to test a source method named \lstInline{daysBetween}.
In addition, the name of the class that this test method belongs to is 
\lstInline{TestDays}, which indicates that the tests in this class are related
to the \lstInline{Days} class in the source code. Based on this observation, we
build new input features for the test method, and their target source method, by
concatenating their method and class names, respectively. To handle these
features, we use a Deep Neural Network (DNN) model consisting of the word embedding
layers and GRUs~\cite{cho2014learning}. See Section~\ref{sec:model_architecture}
for more details of our model architecture.

Note that such a linguistic link may not always exist. For example, consider one
of the test methods in commons-lang named \lstInline{testLang865}, which is
specifically designed to handle the bug report whose unique identification number is 865.\footnote{\url{https://issues.apache.org/jira/projects/LANG/issues/LANG-865}}
The test actually checks whether \lstInline{LocaleUtils.toLocale} can parse 
strings starting with an underscore, but there is no clue of this goal in the 
name of the test method.

\subsubsection{Code Tokens of Mutated Statement}

Given a source and a test method, their names as features remain identical for
all the mutants that are generated within the source method. However, they do
not survive, or get killed by, the same test method collectively. We need
additional features that allow us to distinguish individual mutants. To capture
the characteristic of each mutant, we take the code snippet of the line in which
the mutant is generated, as well as the actual token(s) before and after the
mutation. Note that this triplet of information is often provided by the
mutation tool themselves.\footnote{Since PIT does not provide before and
after tokens in the report, we exclude them when training models for PIT.}

For example, consider the source method in Figure~\ref{fig:mutated_stmt_ex}.
Major mutation tool mutates Line 2 by changing \lstinline+array.length == 0+ to
\lstinline+array.length >= 0+. We use the entire if-statement as well as
before/after code fragments:

\begin{itemize}
    \item Mutated Statement: \lstinline+if (array == null || array.length == 0)+
    \item Before: \lstinline+array.length == 0+
    \item After: \lstinline+array.length >= 0+
\end{itemize}

Similar to the method name features, code tokens of mutated statements are
processed by the DNN model. In particular, the before and after code tokens are
compared to each other using comparison layers (see
Section~\ref{sec:model_architecture} for more details).

\subsubsection{Mutation Operator} 
Lastly, we use mutation operator as a categorical input feature to explicitly 
represent how destructive the operator is. We posit that it may be easier to
kill the mutants generated using Return Values mutator than Binary Arithmetic
Operation mutator. We use one-hot encoding to represent the type of mutation
operator. Note that the dimension of the one-hot encoding depends on
the number of mutation operators supported by the mutation tool. PIT provides 
11 operators in its default configuration and Major provides nine operators in 
our setting.

\subsection{Preprocessing}
All textual inputs go through the following preprocessing.

\begin{itemize}
\item Word Filtering: The numeric or string literals may exhibit local features
that would be difficult to generalise. To avoid overfitting to the local
features, we filter numeric and string literals by replacing them with special
tokens. Also, we use two more special tokens for unknown word and for removed
words due to the mutation.

\item Subword Splitting: Compound words tend to convey several concepts. For
instance, \lstInline{ConvertToAUTF8String} can be seen as a compound of  
`Convert', `To', `A', `UTF8', and `String'. The compound words not only 
increase the vocabulary size of the corpus, but also present challenges for 
effective learning of the word embedding due to their rareness~\cite{
Karampatsis2020ICSE}. To address this issue, we employ the state-of-the-art subword splitter Spiral~\cite{hucka2018spiral} and segment the compound words into subwords. 
\end{itemize}

Finally, all preprocessed tokens for all input features are aggregated to 
build a dataset and vocabulary. Each element in the dataset represents 
an one-to-one mapping between a mutant and a test, labelled 0 if the mutant 
survives the test, and 1 if killed. The dataset is made up of only tests that 
cover the mutant: label 0 means that the mutant is covered but not killed by 
the test. This will reduce training and prediction time, as collecting 
coverage is relatively inexpensive compared to the actual mutation analysis. In
the study, we only consider cross-version scenarios, and not cross-project scenarios.
\new{This means that we will be less affected by the Out-of-Vocabulary (OOV) problem~\cite{hellendoorn2017deep}.}

\subsection{Model Architecture}
\label{sec:model_architecture}

\begin{figure}[!ht]
    \centering
    \includegraphics[width=0.65\textwidth]{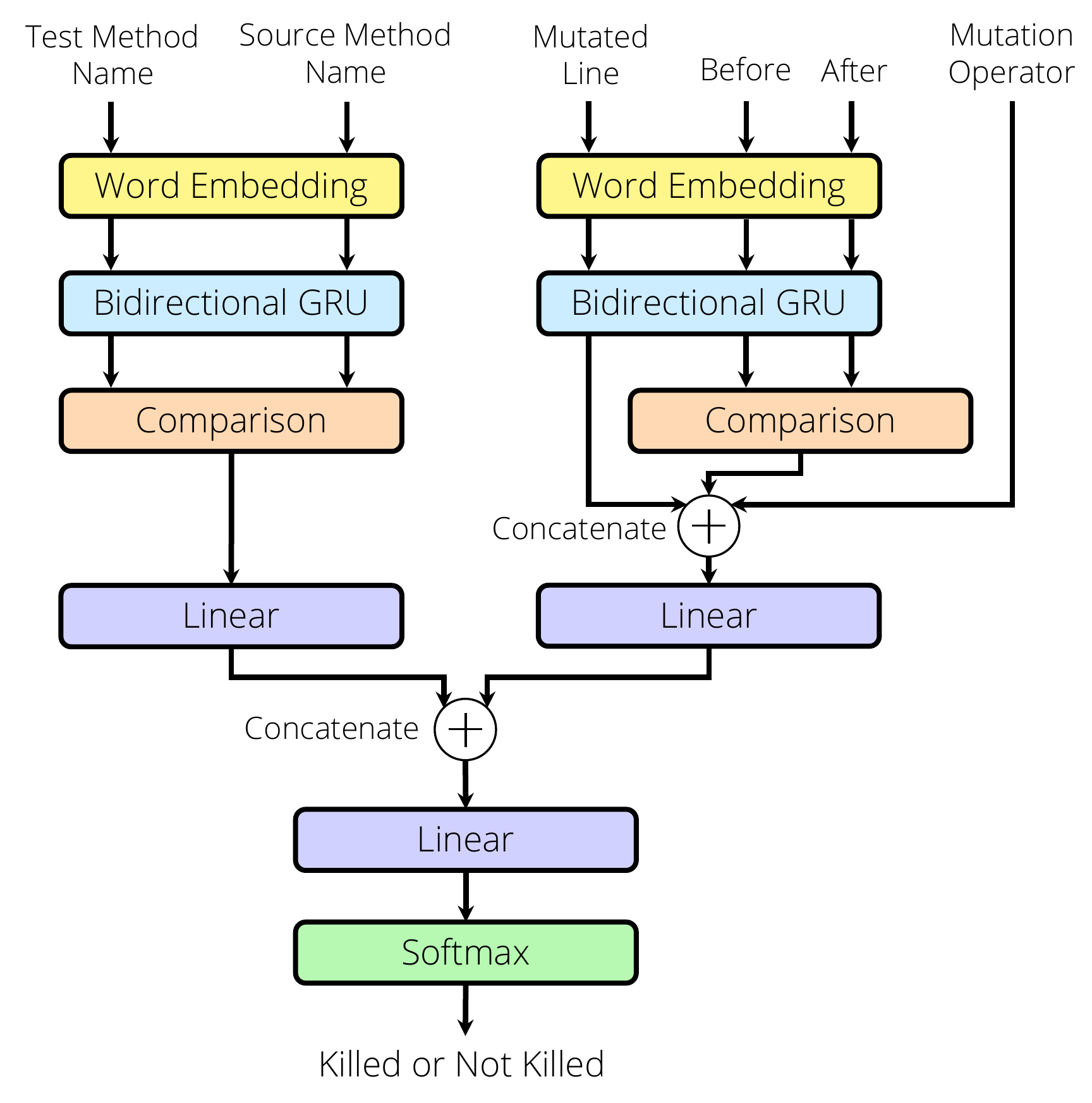}
    \caption{Model architecture of \name}
    \label{fig:model_architecture}
\end{figure}

Figure~\ref{fig:model_architecture} illustrates a model architecture of \name.
It consists of the word embedding layers to convert words to vector
representations, the bidirectional GRU layers to extract sequential context in names
and code tokens, the comparison layers to quantify differences between two vector
representations, and the linear layer and softmax for the final classification. We
detail each component in the following subsections.

\subsubsection{Input Layer}

The input layer consists of a name-based and a mutation-specific part.
We argue that the name-based input features mostly reflect the NL channel in 
the code, whereas the mutation-specific input features reflect more of 
the AL channel in the code, as the code tokens contain some parts of program 
logic. Consequently, we use independent word embedding layer, as well as the 
bidirectional GRU, for each part.

\subsubsection{Word Embedding and Encoding Layer}

The word embedding layer maps each word to a numerical representation that captures
the relative relationship between words. We use $\mathbf{E_p} \in
\mathbb{R}^{|V_p| \times d}$ for the name-based features, and $\mathbf{E_q} \in
\mathbb{R}^{|V_q| \times d}$ for the mutant-specific features: $V_p$ and $V_q$
denote vocabularies of the names and code tokens respectively, and $d$ denotes a
dimension of the word embedding. We train the embedding layer from scratch
as part of the model training, instead of using pre-trained weights. 

The words in the test method name $\{w_{t,1}, \dots , w_{t,n}\}$ and the source
method name $\{w_{s,1}, \dots , w_{s,m}\}$ are passed to the word embedding 
layer and converted to $x_{t,i} = \mathbf{E_p}(w_{t,i}) \in \mathbb{R}^{d}$, 
$1 \leq i \leq n$ and $x_{s, j} = \mathbf{E_p}(w_{s, j}) \in \mathbb{R}^{d}$, 
$1 \leq j \leq m$. Subsequently, the bidirectional GRU is used to extract hidden 
context between words in two directions. For the test method name:

\begin{equation}
    \overrightarrow{h_{t,i}} =  \overrightarrow{\mathbf{GRU}}(x_{t,i})
\end{equation}
\begin{equation}
    \overleftarrow{h_{t,i}} =  \overleftarrow{\mathbf{GRU}}(x_{t,i})
\end{equation}
\begin{equation}
    h_{t,i} = \overrightarrow{h_{t,i}} \oplus \overleftarrow{h_{t,i}}
\end{equation}

The hidden representations of a forward GRU and a backward GRU are concatenated,
composing one representations $h_{t,i}$. Next, we adopt an attention mechanism to
reward those words that are deemed to be important.

\begin{equation}
    u_{t,i} =\tanh \left(W_{att} h_{t,i} + b_{att}\right)
\end{equation}
\begin{equation}
    \alpha_{t,i} =\frac{\exp \left(u_{t,i}^{\top} u\right)}{\sum_{i} \exp \left(u_{t,i}^{\top} u\right)}
\label{eq:attention_softmax}
\end{equation}
\begin{equation}
    v_t =\sum_{i} \alpha_{t,i} h_{t,i}
\label{eq:weighted_sum}
\end{equation}

$W_{att}$ denotes a learnable weight matrix and $b_{att}$ is a corresponding
bias. The attention vector is normalised by softmax
(Equation~\ref{eq:attention_softmax}) and makes resulting embedding vector $v_t$
through the weighted sum with $h_{t,i}$ (Equation~\ref{eq:weighted_sum}). The same
bidirectional GRU and attention mechanism are applied to $h_{s,j}$, computing
$v_s$. As a result, the $v_t$ and $v_s$ are embedding vectors of test and source
method name, respectively, after the bidirectional GRU layer.

The mutation-specific part of the input uses an independent embedding layer 
$\mathbf{E_q}$ and a bidirectional GRU layer. Except the mutation operator feature 
that is categorical data and thus one-hot encoded, mutation-specific features 
are fed into $\mathbf{E_q}$, the bidirectional GRU, and attention mechanism in the 
same way as the name-based features. At the end of this process, we get three 
embedding vectors: $v_l$ (mutated line), $v_b$ (before), and $v_a$ (after).

\subsubsection{Comparison Layer}

\name is based on the intuition that the semantic similarity between the names
of the source and test method will reflect, and eventually allow us to learn,
the mutant-test relationship. To capture the semantic similarity between
names, we use a comparison layer~\cite{wang2016compare, hoang2020cc2vec} to
measure similarity between two embedding vectors, $v_t$ and $v_s$. The
comparison layer includes various comparison functions: Neural Tensor Network
(Bilinear layer), Neural Network (Linear layer), Cosine and Euclidean
similarity, element-wise subtraction and multiplication. For example, Neural
Network is a simple network with one linear layer and Neural Tensor Network is
similar but with bilinear layer, each of which has learnable parameters
$W_{NN}$, $b_{NN}$, $W_{NT}$, $b_{NT}$:

\begin{equation}
    c_{NN}=ReLU(W_{NN}(v_t \oplus v_s) + b_{NN})
\end{equation}
\begin{equation}
    c_{NT}=ReLU(v_{t}^{T}W_{NT}v_{s} + b_{NT})
\end{equation}
Other comparison functions do not have learnable parameters, and simply compute
distances between two input vectors: $c_{cos}$ (cosine
similarity), $c_{euc}$ (Euclidean distance), $c_{sub}$ (element-wise
subtraction), and $c_{mul}$ (element-wise multiplication). The comparison
vectors are concatenated and form an embedding vector $v_{ts}$:
\begin{equation}
    v_{ts} = c_{NN} \oplus c_{NT} \oplus  c_{cos} \oplus c_{euc} \oplus c_{sub} \oplus c_{mul}
\end{equation}

With the mutant-specific inputs, before and after represent the changes of the same
part of the code. Therefore, we apply the comparison functions to the embedding
vector of the before ($v_{b}$) and after ($v_{a}$): $v_{ba}$. Note that some
comparison functions such as Neural Tensor Network contain learnable parameters,
we use separate comparison functions for before and after. To group and reduce
the dimensionality of each embedding vector, $v_{ts}$ and $v_{ba}$, we use two linear
layers and concatenate them:

\begin{equation}
    v_{tsba} = W_{ts}(v_{ts}) \oplus W_{ba}(v_{ba})
\end{equation}

The embedding vector $v_{tsba}$ is passed to the final linear layer and the 
softmax function to produce the probability of the input belonging to each 
class (i.e., killed or not killed).

\begin{equation}
    p = softmax(W(v_{tsba}))
\end{equation}

During training, all learnable weights such as the word embeddings, the weight matrices, or the biases are
learnt by minimizing the cross entropy loss.

\section{Experimental Setup}
\label{sec:experimental_setup}

We design an empirical study that evaluates \name on seven different Java
projects consisting of 37 program versions in \dfj.

\subsection{Research Question}
We ask the following six research questions to evaluate \name:

\subsubsection{RQ1. Effectiveness}
\new{\emph{Can \name learn and predict a kill matrix using the NL channel in the
source code and tests? How does \name perform against PMT and a coverage based
heuristic?} RQ1 concerns the effectiveness of \name under a cross-version
scenario with the comparisons to other models. We answer RQ1 by training \name
using the actual kill matrices of earlier versions of a subject system, and
using it to predict the kill matrices of subsequent versions. We also train a
test case level PMT and a coverage based baseline to compare the
F-score for each subject.}

\subsubsection{RQ2. Efficiency} 

\emph{How efficient is \name compared to actually performing mutation analysis
to obtain an entire kill matrix?} RQ2 concerns how much execution time can be
saved by \name, compared to the execution time of the traditional mutation
analysis. We report the time required for preprocessing and predicting the full
kill matrix.

\subsubsection{RQ3. Generalisation} 

\emph{How well does \name generalise to test cases that are newly added to a 
subject system?} If \name truly learns through the NL channel, it should retain its predictive power for the new 
and unseen test cases, as well as the unseen mutants. If \name simply memorises the 
features, on the other hand, its prediction accuracy for 
newly introduced test cases will be dramatically lower than that for existing 
test cases. We answer RQ3 by evaluating \name separately, for the old and new test 
cases.

\subsubsection{RQ4. Mutation Score} 

\emph{How well does \name predict whether a mutant is killed or survives by
a given test suite (i.e., mutation score) compared to PMT?} RQ4 concerns whether
the full kill matrix predicted by \name actually produces an accurate aggregated
result over the test suite as well. If it does, the results would support that
\name incorporates the coarser grained PMT. We answer RQ4 by computing the
mutation score and F-score using predicted kill matrix by \name, and comparing them
to the actual mutation score and F-score of PMT.

\subsubsection{RQ5. Application Study} 

\add{
\emph{Can predicted kill matrix by \name be used for mutation based fault localisation?} 
Beyond the mutation score, RQ5 concerns whether \name can be applied for
Mutation Based Fault Localisation (MBFL) techniques that use the kill matrix to
locate faults. To this end, we apply \name to SIMFL~\cite{kim2021issre}, a
state-of-the-art MBFL technique that relies on statistical inference over the
kill matrix (See Section~\ref{sec:mbfl} for the details of SIMFL). Using the
predicted kill matrix by \name, we attempt to localise faults using the predicted
kill matrices, without paying the cost of mutation analysis after the bug has
been detected. Since we use the predicted kill matrices instead of real ones, we
expect the localisation results to be less accurate. Thus, we report the changes
of localisation accuracy, and also compare their effectiveness to two widely
studied MBFL techniques, MUSE~\cite{Moon:2014ly} and
Metallaxis~\cite{Papadakis:2015sf}.}

\subsubsection{RQ6. Naming Convention and Automated Test Generation}
\add{
\emph{How sensitive is \name when applied to automatically generated test cases 
and their naming convention?} With RQ6, we evaluate \name using a test data 
automatically generated by EvoSuite~\cite{Fraser:2013vn}. We begin by training 
\name with a kill matrix of EvoSuite generated test suite, and predicting 
the kill matrix of another, independently generated EvoSuite test suite for a 
subsequent version. In addition, we also cross-evaluate the performance of \name
between the developer written test suites and the EvoSuite generated test suites. The 
use of the machine generated test cases has multiple implications. First, unlike 
the developer written test suites that gradually evolve with the SUT, EvoSuite 
test suites are generated for each version in our experimental protocol, 
getting rid of the continuity in the test suite contents. Second, the machine generated 
test cases follow different naming conventions from those used by developers. 
We adopt the descriptive naming strategy implemented in 
EvoSuite~\cite{Daka2017zf} and see how much information can be obtained from 
the NL channel. Finally, we expect the test adequacy of the machine generated test 
suites to be different from those written by developers in terms of the mutation 
testing. In the evaluation, we investigate how much the predictive power of \name is 
affected by these factors when we replace the developer written tests with the 
automatically generated tests using EvoSuite.}

\subsection{Subject Program}
\label{sec:subject_program}

We select 37 subject program versions on seven different projects in \dfj v2.0.0, which
has been widely used in the software testing research, and provides an infrastructure
that makes tests run, coverage analysis, and mutation analysis to be easily
performed and reproduced. Table~\ref{table:subject} lists the latest and the
oldest subject program versions for each project. Column 1 shows project name,
and the number of program versions in each project we use in the study. We
denote the version number by the identifier number used in \dfj, and denote the
two adjacent versions, or a version and immediately preceding version by the two
versions whose version numbers are right next or before to each other.

To facilitate cross-version scenario, we select programs whose version number
is a multiple of five or ten. Note that smaller version number does not
always represent more recent codebase; Gson, Cli, JC, and Csv assign smaller
version number to older codebase. Chart is the largest subject with 96k LoC and
Time has 4k tests which is the most and 74 times larger than tests of Csv which
contains only 54 tests. We exclude subjects of Time under the Major mutation tool
because their mutation analyses for entire kill matrix have not been completed
within 48 hours.

\begin{table}[!t]
    \centering
    \caption{Subject Program}
    \label{table:subject}
    \scalebox{0.75}{
        \begin{tabular}{l|ll|rrr|rr|rr}
            \toprule            
            Project & Identifier & Version & LoC   & \# Tests & Date & \multicolumn{2}{c|}{Major} & \multicolumn{2}{c}{PIT}        \\
              &     &     &       &          &  &  \# Mut. Gen.  &  Killed \% & \# Mut. Gen.  & Killed \%   \\
            \midrule
            commons-lang & Lang & 1 & 21,788 & 2,291 & 2013-07-26 & 22,793  & 74.2\% & 10,546  & 85.4\%  \\
              & Lang & 10 & 20,433 & 2,198 & 2012-09-27 & 19,767  & 74.7\% & 9,477  & 85.6\%  \\
              & Lang & 20 & 18,967 & 1,876 & 2011-07-03 & 19,073  & 74.5\% & 8,994  & 84.3\%  \\
              & Lang & 30 & 17,660 & 1,733 & 2010-03-16 & 18,144  & 74.7\% & 8,234  & 85.5\%  \\
              & Lang & 40 & 17,435 & 1,643 & 2009-10-22 & 17,972  & 74.1\% & 8,138  & 73.8\%  \\
              & Lang & 50 & 17,760 & 1,720 & 2007-10-31 & 18,151  & 73.0\% & 8,744  & 81.3\%  \\
              & Lang & 60 & 16,920 & 1,590 & 2006-10-31 & 17,819  & 72.6\% & 8,506  & 81.5\%  \\
            \midrule
            joda-time & Time & 1 & 27,801 & 4,041 & 2013-12-02 & 20,257  & 65.1\% & 9,706  & 78.4\%  \\
              & Time & 5 & 27,664 & 4,013 & 2013-11-01 & 20,144  & 65.1\% & 9,658  & 78.6\%  \\
              & Time & 10 & 27,341 & 3,954 & 2013-06-16 & 19,991  & 65.2\% & 9,634  & 78.0\%  \\
              & Time & 15 & 27,215 & 3,894 & 2012-04-30 & 19,756  & 65.3\% & 9,604  & 77.9\%  \\
              & Time & 20 & 27,156 & 3,868 & 2011-10-23 & 19,618  & 65.1\% & 9,556  & 77.9\%  \\
              & Time & 25 & 26,805 & 3,810 & 2010-12-05 & 19,488  & 65.5\% & 9,416  & 77.7\%  \\
            \midrule
            jfreechart & Chart & 1 & 96,382 & 2,193 & 2010-02-09 & 81,006  & 23.6\% & 35,690  & 33.9\%  \\
              & Chart & 5 & 89,347 & 2,033 & 2008-11-24 & 75,024  & 23.6\% & 33,157  & 33.7\%  \\
              & Chart & 10 & 84,482 & 1,805 & 2008-06-10 & 71,052  & 22.8\% & 31,183  & 33.1\%  \\
              & Chart & 15 & 84,134 & 1,782 & 2008-03-19 & 70,647  & 22.6\% & 30,977  & 32.8\%  \\
              & Chart & 20 & 80,508 & 1,651 & 2007-10-08 & 67,479  & 22.2\% & 30,471  & 31.4\%  \\
              & Chart & 25 & 79,823 & 1,617 & 2007-08-28 & 66,766  & 22.1\% & 30,213  & 31.0\%  \\
            \midrule
            gson & Gson & 15 & 7,826 & 1,029 & 2017-05-31 & 5,044  & 64.2\% & 2,670  & 76.1\%  \\
              & Gson & 10 & 7,693 & 996 & 2016-05-17 & 4,775  & 65.4\% & 2,576  & 75.7\%  \\
              & Gson & 5 & 7,630 & 984 & 2016-02-02 & 4,722  & 64.6\% & 2,546  & 75.8\%  \\
              & Gson & 1 & 5,418 & 720 & 2010-11-02 & 2,295  & 61.8\% & 1,564  & 70.3\%  \\
            \midrule
            commons-cli & Cli & 30 & 2,497 & 354 & 2010-06-17 & 1,592  & 81.3\% & 710  & 89.2\%  \\
              & Cli & 20 & 1,989 & 148 & 2008-07-28 & 1,118  & 77.3\% & 509  & 83.5\%  \\
              & Cli & 10 & 2,002 & 112 & 2008-05-29 & 1,151  & 68.4\% & 515  & 77.9\%  \\
              & Cli & 1 & 1,937 & 94 & 2007-05-15 & 1,118  & 66.3\% & 499  & 73.3\%  \\
            \midrule
            jackson-core & JC & 25 & 25,218 & 573 & 2019-01-16 & 30,010  & 53.8\% & 13,198  & 63.5\%  \\
              & JC & 20 & 21,480 & 384 & 2016-09-01 & 25,257  & 49.8\% & 11,115  & 58.5\%  \\
              & JC & 15 & 18,652 & 346 & 2016-03-21 & 21,599  & 48.1\% & 9,632  & 56.1\%  \\
              & JC & 10 & 18,930 & 330 & 2015-07-31 & 22,089  & 48.6\% & 9,773  & 56.6\%  \\
              & JC & 5 & 15,687 & 240 & 2014-12-07 & 18,610  & 47.3\% & 8,227  & 54.6\%  \\
              & JC & 1 & 15,882 & 206 & 2013-08-28 & 16,982  & 46.5\% & 7,561  & 53.0\%  \\
            \midrule
            commons-csv & Csv & 15 & 1,619 & 290 & 2017-12-11 & 1,173  & 71.4\% & 599  & 84.0\%  \\
              & Csv & 10 & 1,276 & 200 & 2014-06-09 & 1,043  & 71.1\% & 493  & 81.9\%  \\
              & Csv & 5 & 1,236 & 189 & 2014-03-13 & 996  & 72.3\% & 469  & 81.9\%  \\
              & Csv & 1 & 806 & 54 & 2012-03-27 & 695  & 68.1\% & 280  & 80.4\%  \\
            
            \bottomrule
        \end{tabular}
    }
\end{table}

\subsection{Mutation Tool}
Since the result of mutation analysis is highly dependent on the mutation tool
and its configuration, we evaluate \name using two widely studied Java mutation
tools, PIT ver. 1.5.2~\cite{Coles2016ft} and Major ver. 
1.3.4~\cite{just2014major}. 
PIT uses bytecode mutator that is known to be efficient and well integrated 
with various development environments. Meanwhile, Major adopts 
compiler-integrated mutator that transforms the abstract syntax tree (AST) and 
provides in-depth mutation report. Table~\ref{table:subject} lists the number 
of generated mutants, killed mutants, and kill percentage of PIT and Major. It 
shows that large LoC leads to the large number of mutants: Major generates up 
to 81k mutants for Chart whereas it generates 11k mutants for Csv.

In the study, we filter mutants of PIT that are not captured by our preprocess
steps, or killed due to implicit oracles such as time out or uncaught out of
memory exceptions, because PIT does not report the
tests that caused the kill. Therefore, in Section~\ref{sec:rq4}, we have
computed the mutation score ourselves after excluding all those mutants, not
using the score reported by the mutation tool.

\subsection{PMT and Coverage based Baseline Model}
\new{To compare the performance of \name against other models, we choose two
models: PMT and a simple coverage based baseline model. Following the
recent study that has investigated the model choices for PMT, we implement a
Random Forest classifier using 12 features that has shown the best
performance~\cite{Mao2019ur}. Most of the features of PMT are collected in the
same way, but the test suite level features such as \textit{numExecuteCovered}
are collected in the test case level for PMA. In addition, as a sanity check, 
we also include a coverage based model that predicts all mutants covered by any test will be killed. Coverage information for this heuristic has been collected using Cobertura.}

\subsection{Mutation Based Fault Localisation}
\label{sec:mbfl}

\add{Mutation analysis has been successfully exploited for Fault Localisation
(FL), resulting in a family of Mutation Based Fault Localisation (MBFL)
techniques~\cite{Papadakis2019aa}. For instance, using the seeded faults (i.e.,
mutants), MUSE~\cite{Moon:2014ly} computes a suspiciousness score based on the
ratio of fail-becomes-pass tests and pass-becomes-fail tests for each statement.
Metallaxis~\cite{Papadakis:2015sf} observes the mutants that show similar test
results with the faults using Spectrum based Fault Localisation (SBFL)-like
formulas.}

\add{For the application study of RQ5, we apply \name to a state-of-the-art
MBFL technique, SIMFL~\cite{kim2021issre}, a technique that utilises the
entire kill matrices. SIMFL tries to locate faults using a statistical
inference over the mutant-tests results in the kill matrix as follows. First, it computes a
full kill matrix of a given program. Subsequently, SIMFL learns a
statistical model that can predict where the mutant was based on the
patterns of the failed test cases. This model can later be used to localise
faults, as real faults can be considered as yet another mutation that causes
test cases to fail.}

\add{A weakness of SIMFL is that, to use the learnt relationship between mutants
and test cases to localise a new fault, SIMFL should have had access to the
fault revealing test case during its learning phase. However, SIMFL cannot be
applied to a new fault if the fault is revealed by a new test case that was not
part of the original kill matrix it learnt from. \name can help SIMFL overcome
this by inferring the individual mutant-test relationship for the new test case.
With RQ5, we evaluate how much loss in accuracy occurs when we use the inferred kill
matrices.}

\subsection{Evaluation Metric and Protocol}

For all RQs, we use the standard evaluation metric for binary classification, as
PMA is essentially a prediction of binary classes, i.e., killed or not killed.
\add{We compute precision, recall, and F-score, but only report F-score for the
sake of brevity, as they show similar trends. Other metrics and omitted figures
are available from  a supplementary web page at
\url{https://coinse.github.io/seshat-results}. The source code and the full raw
results are publicly available at
\url{https://figshare.com/s/ffbcc010905f3942eaeb}.}

To simulate cross-version scenarios, we set the base versions used for training
the model and apply them to predict the kill matrices of later versions. For
all the two adjacent versions in our subjects, on average, the number of 
elapsed days is 445, the number of added lines is 14,469, and the number of 
removed lines is 14,826.\footnote{We use \texttt{git-diff} tool to extract 
changed lines in \texttt{.java} files.}

\subsection{Configuration and Environment}

We use a default set of mutation operators for PIT and default template file (\lstinline+.mml+)
specified in \dfj for Major. We run mutation analyses using PIT with two threads
and \lstinline+fullMutationMatrix+ options.
For Major, we use \lstinline+sort_methods+ and \lstinline+exportKillMap+
options to enable it to compute the entire kill matrix in a test case level. For the
versions in Chart, we run Major within the \texttt{killmap} tool to reliably
manage stack overflow errors involving JVM crashes.\footnote{We obtained the
tool from
\url{https://bitbucket.org/rjust/fault-localization-data/src/master/killmap}
(commit ea1741dbb) and then modified it to run all given test cases.} 

In the preprocessing step, we perform lexical analysis using a Java lexer implemented in
SLP library.\footnote{\url{https://github.com/SLP-team/SLP-Core}} In training
model, the dimension of word embedding is set to 50, the number of features in
the hidden state of GRU is set to 100, and dropout rate is set to 0.5. The
maximum training epoch is 10.

The mutation analysis using \texttt{killmap} tool was performed 
on four machines each of which runs Ubuntu 18.04.4 LTS on Intel i5-10600 CPU @ 
3.30GHz and 16GB RAM. All remaining experiments were performed on machines 
running Ubuntu 16.04.4 LTS, Intel Xeon E5-2630 v4 CPU, Nvidia TITAN Xp GPU, 
and 256GB RAM.

\begin{figure}[ht]
    \includegraphics[width=0.42\textwidth]{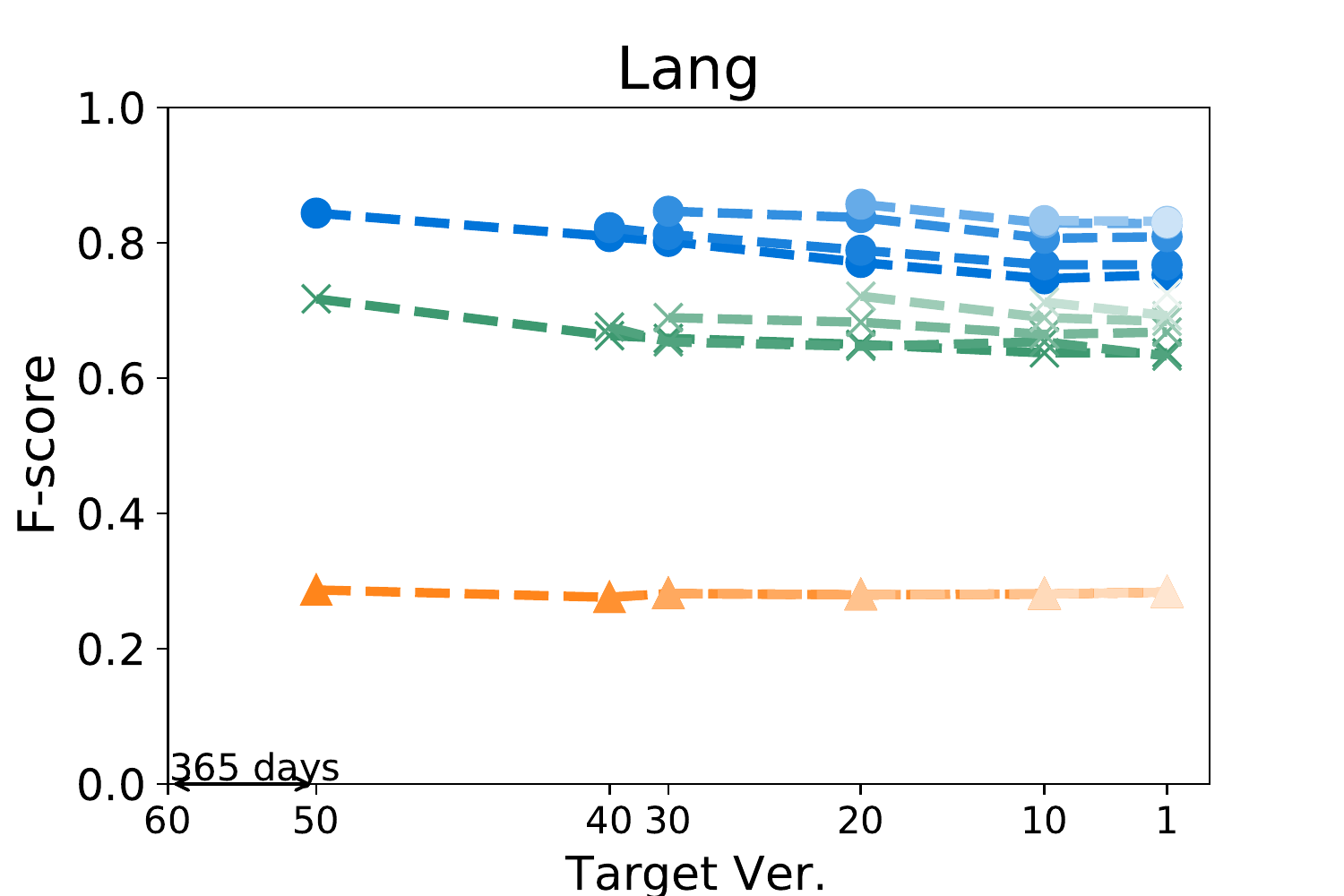}
    \includegraphics[width=0.42\textwidth]{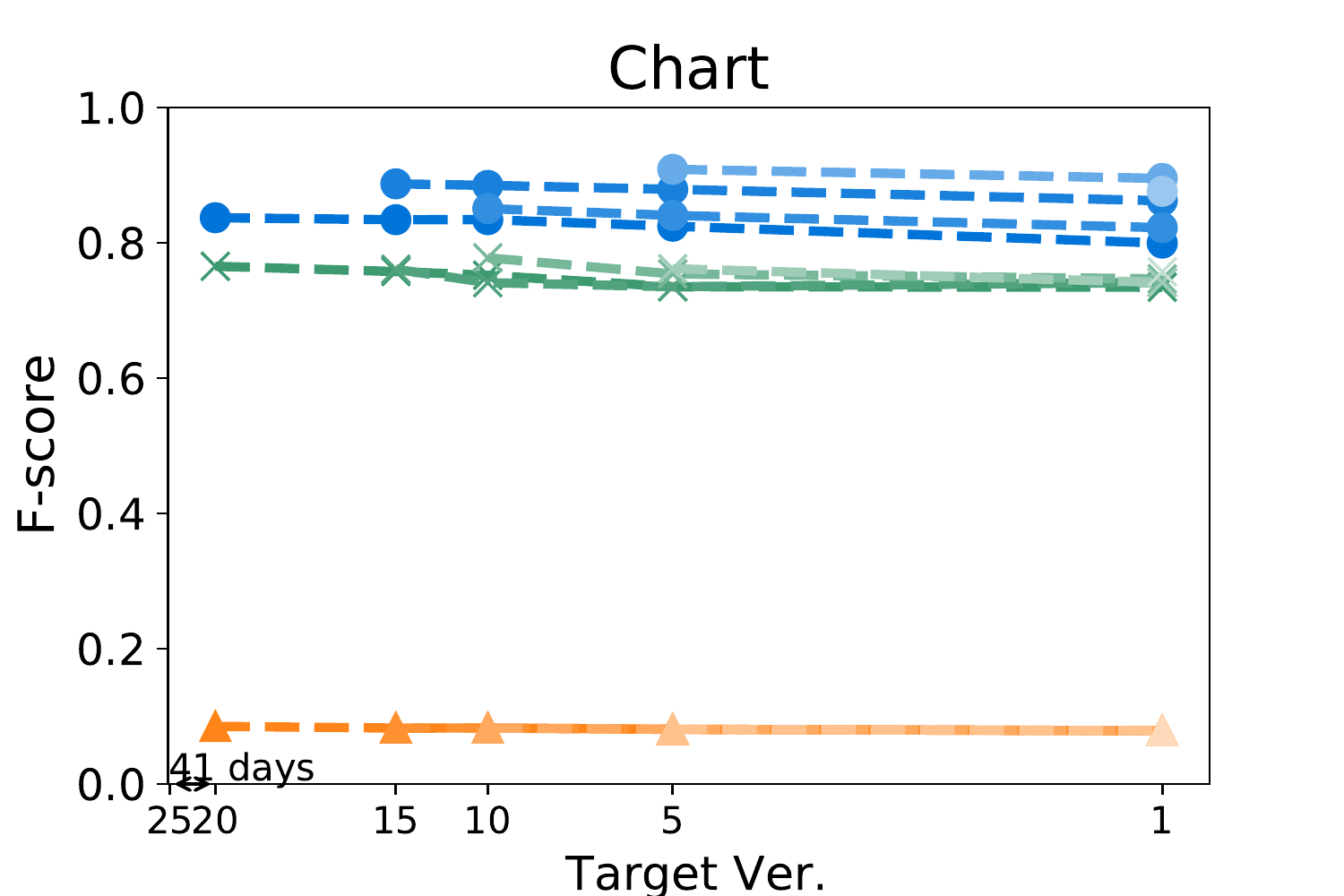}
    \includegraphics[width=0.42\textwidth]{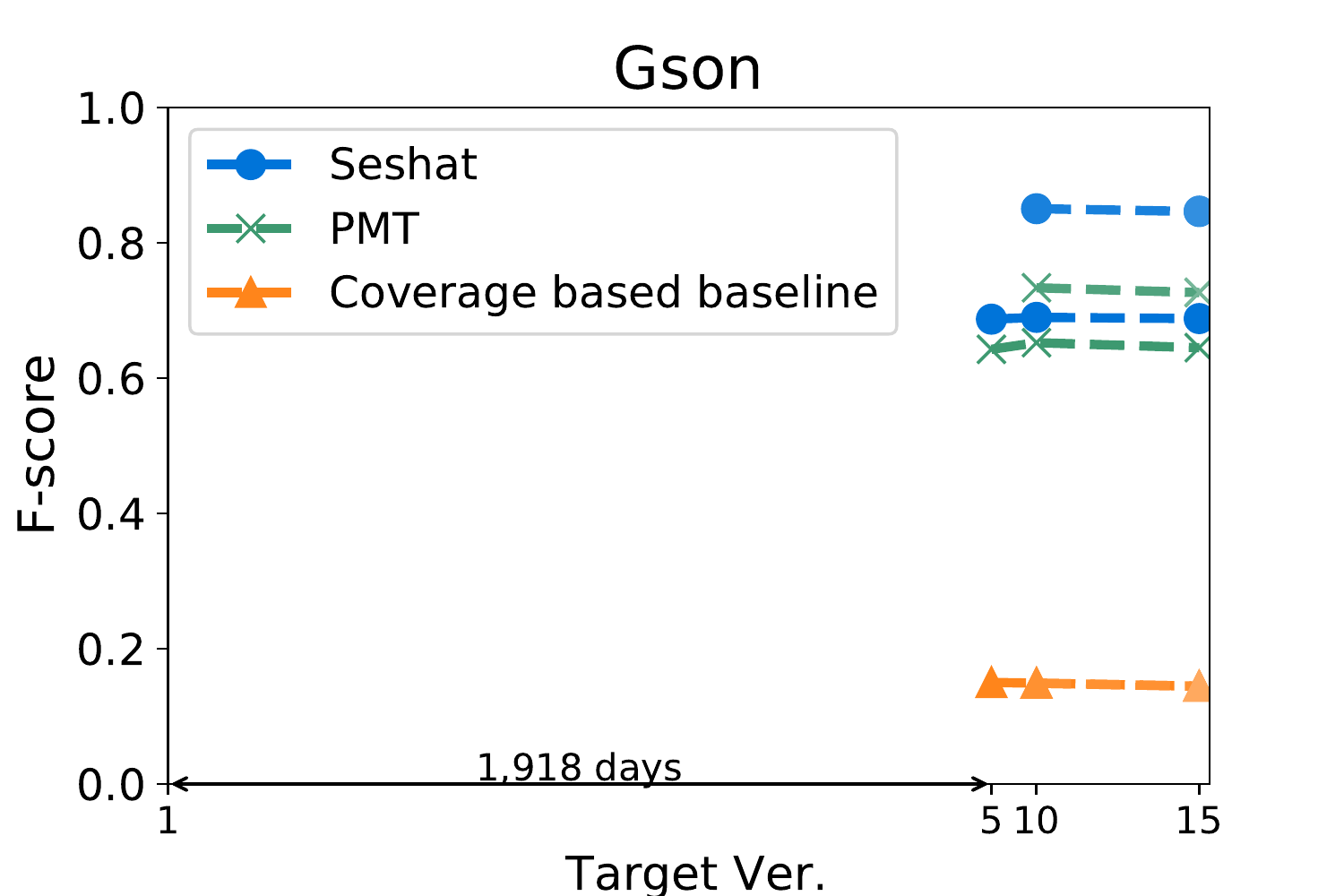}
    \includegraphics[width=0.42\textwidth]{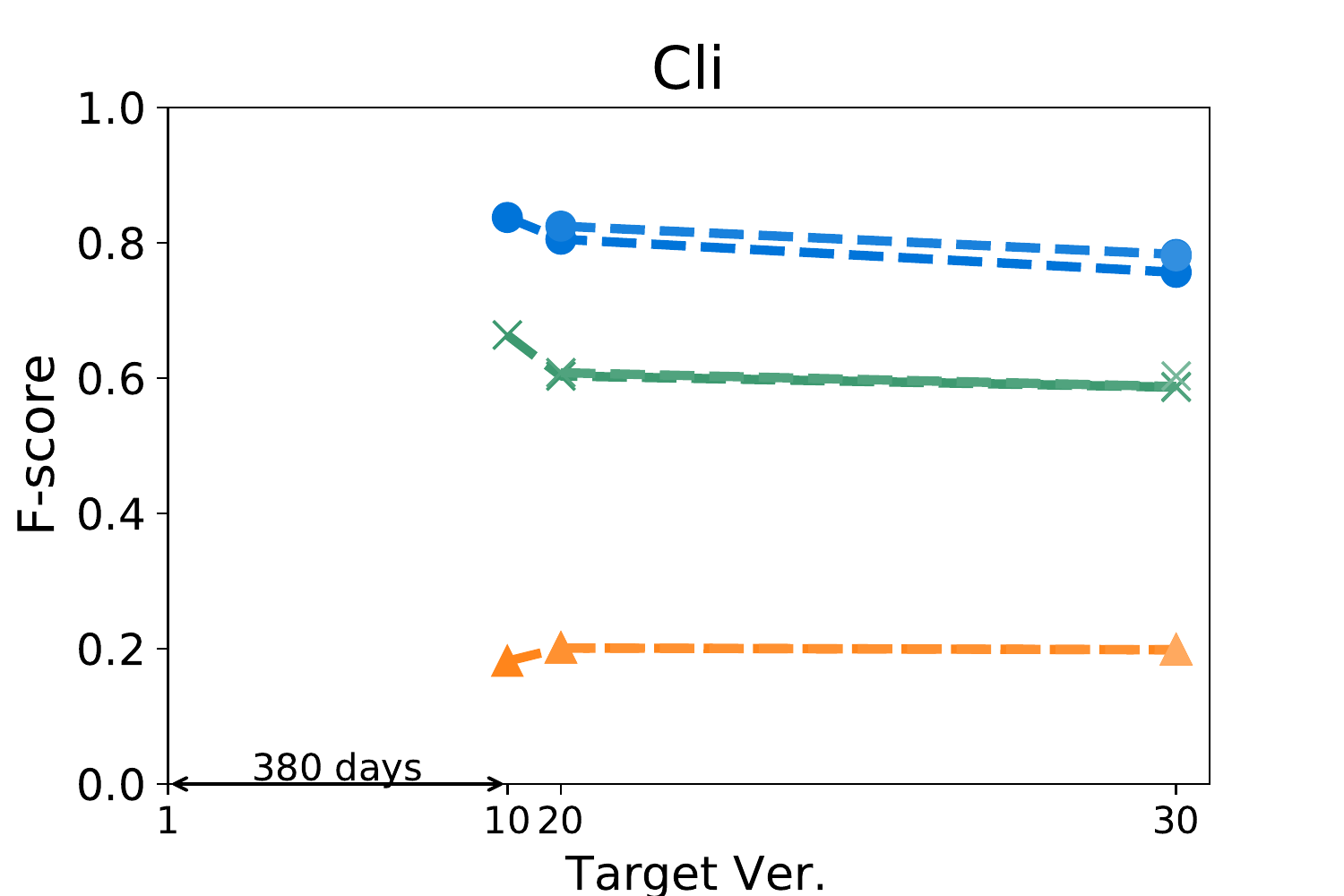}
    \includegraphics[width=0.42\textwidth]{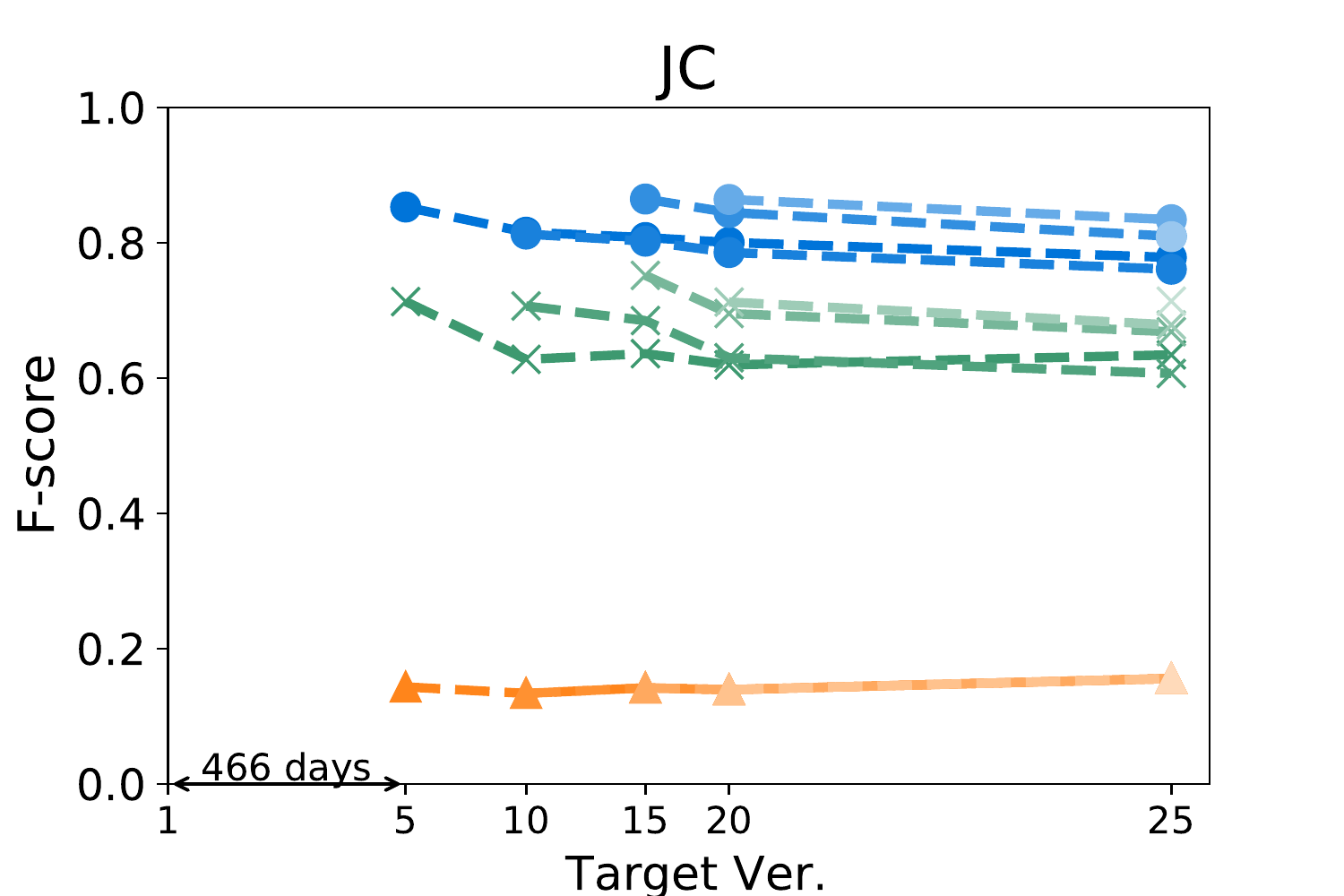}
    \includegraphics[width=0.42\textwidth]{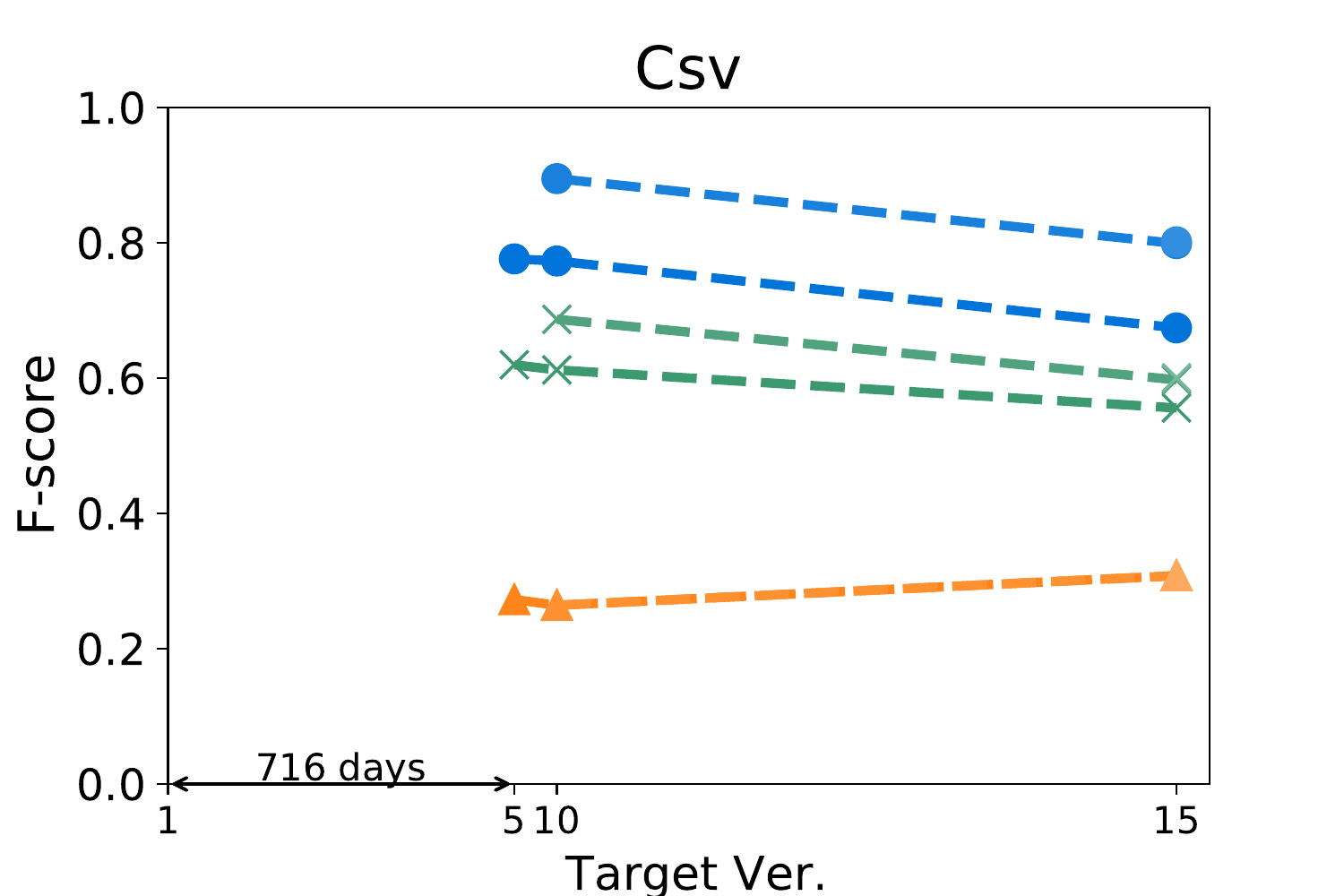}
    \caption{Prediction of the full kill matrix on Major}
    \label{fig:RQ1_major}
    \end{figure}

\begin{figure}[ht]
    \includegraphics[width=0.42\textwidth]{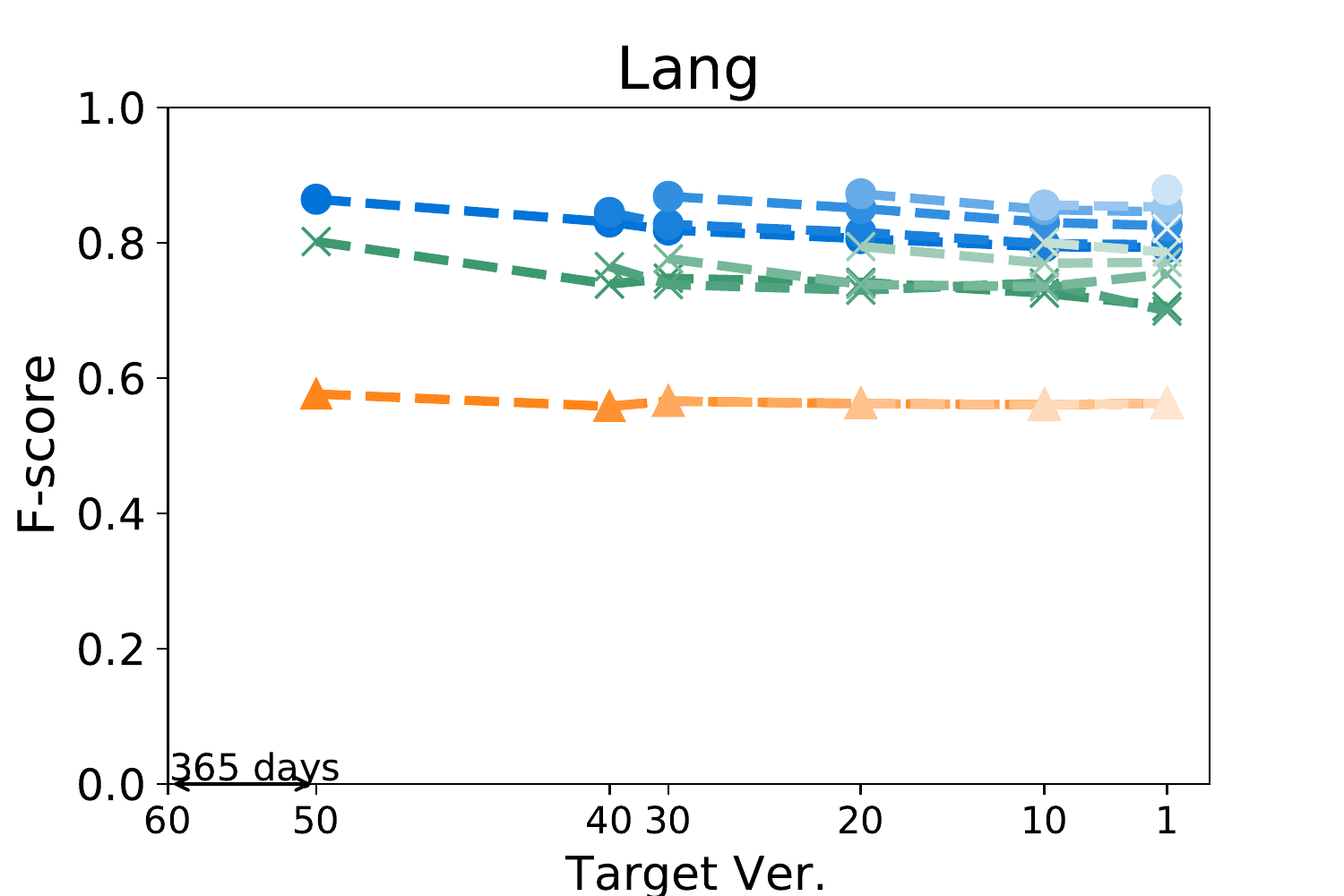}
    \includegraphics[width=0.42\textwidth]{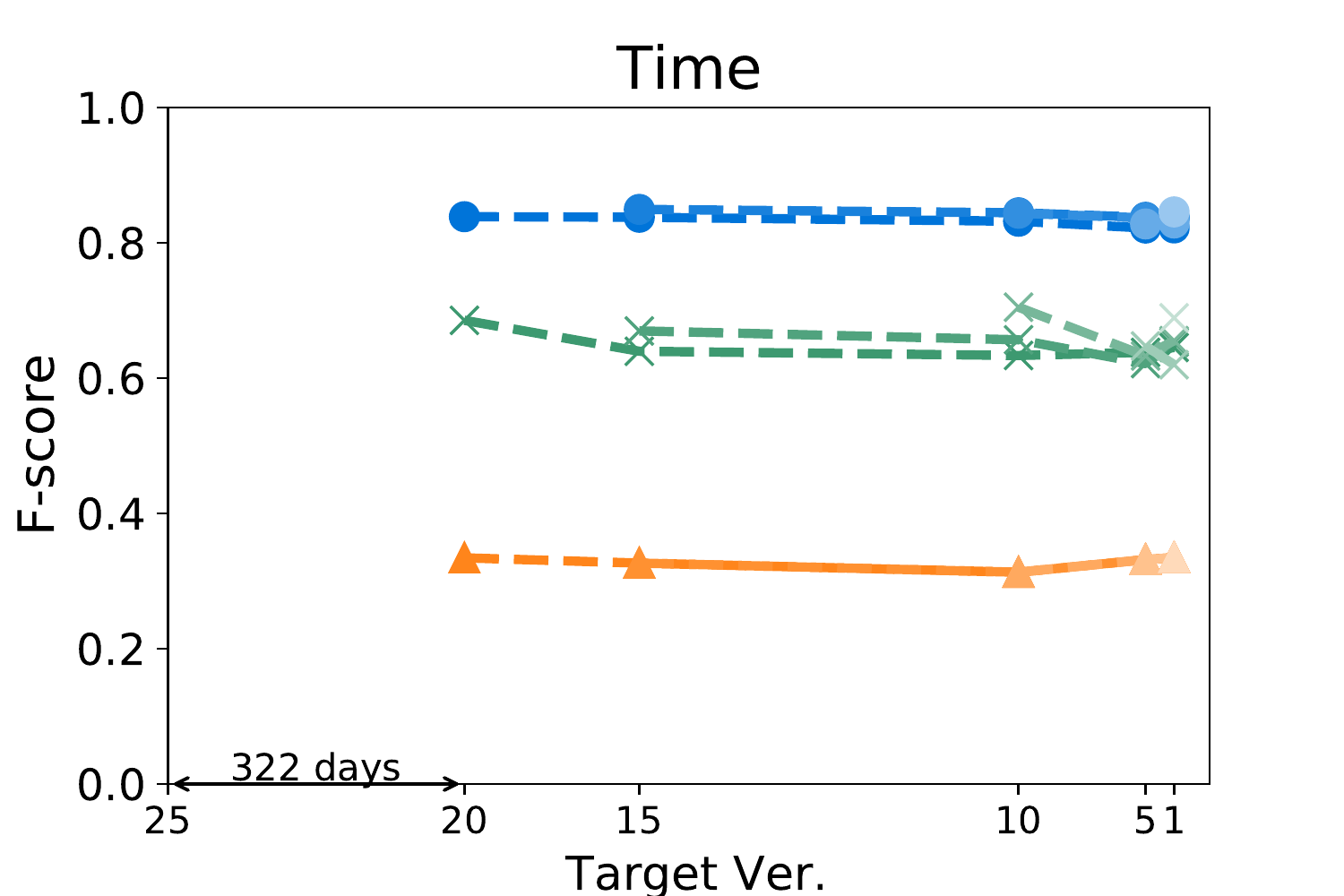}
    \includegraphics[width=0.42\textwidth]{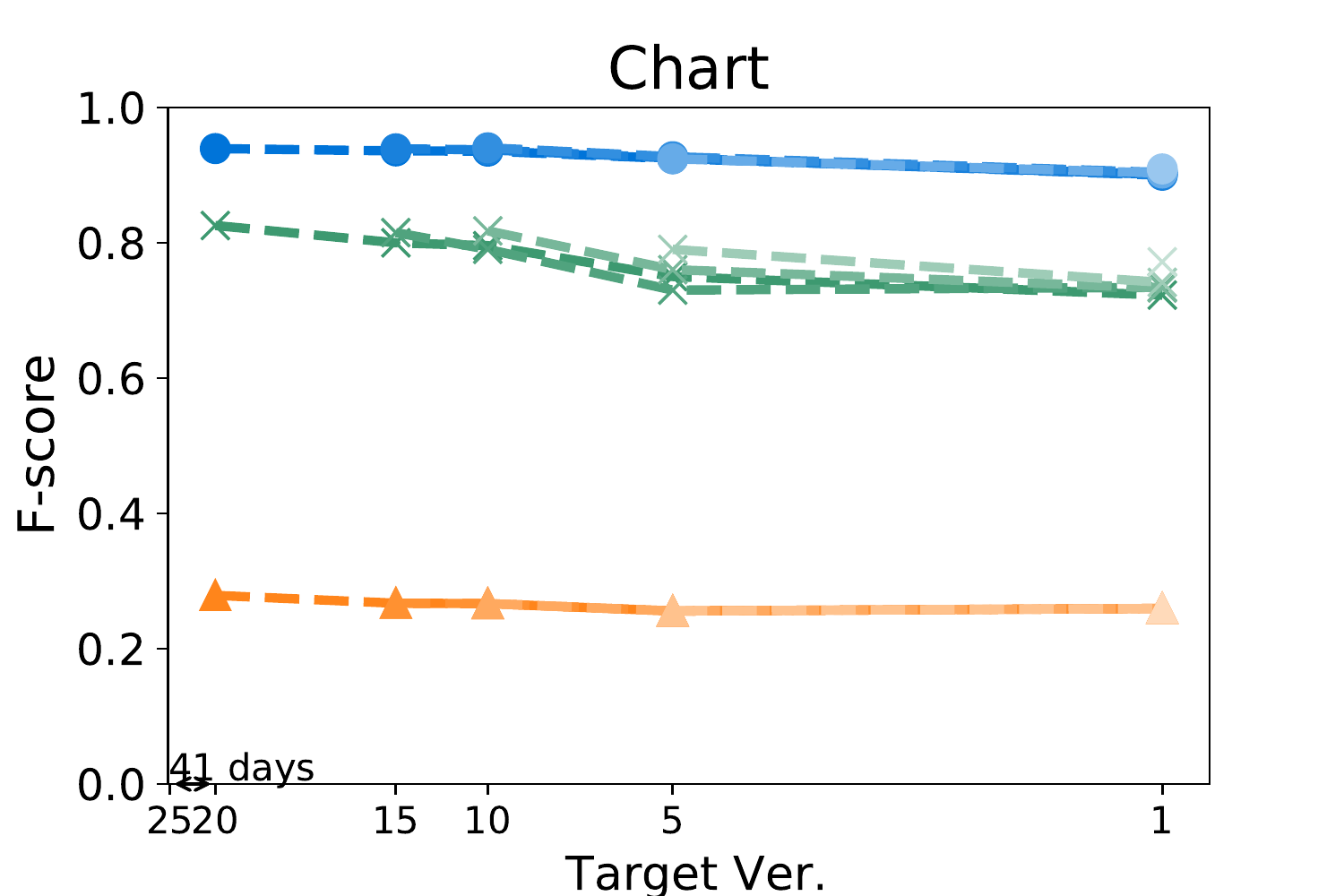}
    \includegraphics[width=0.42\textwidth]{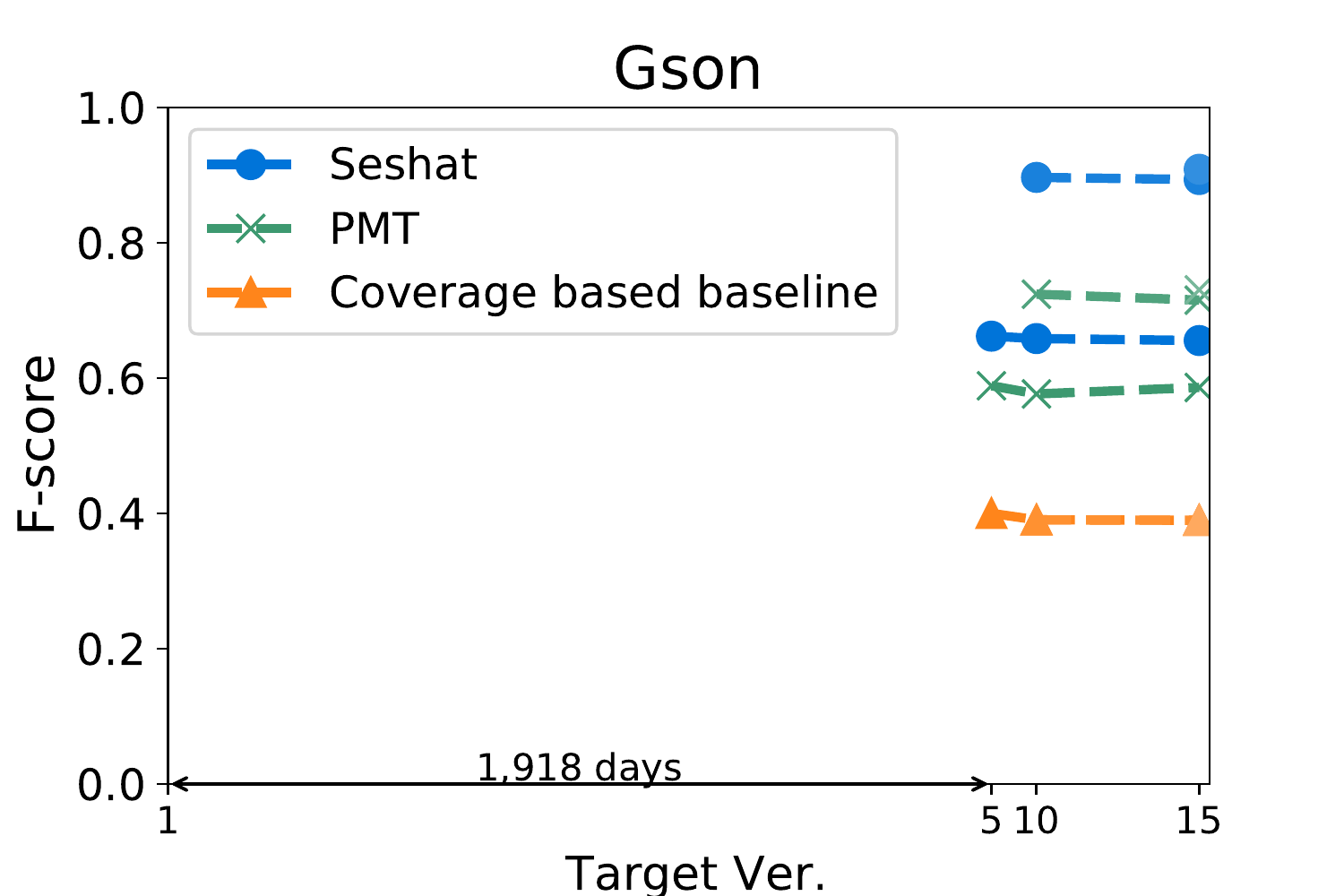}
    \includegraphics[width=0.42\textwidth]{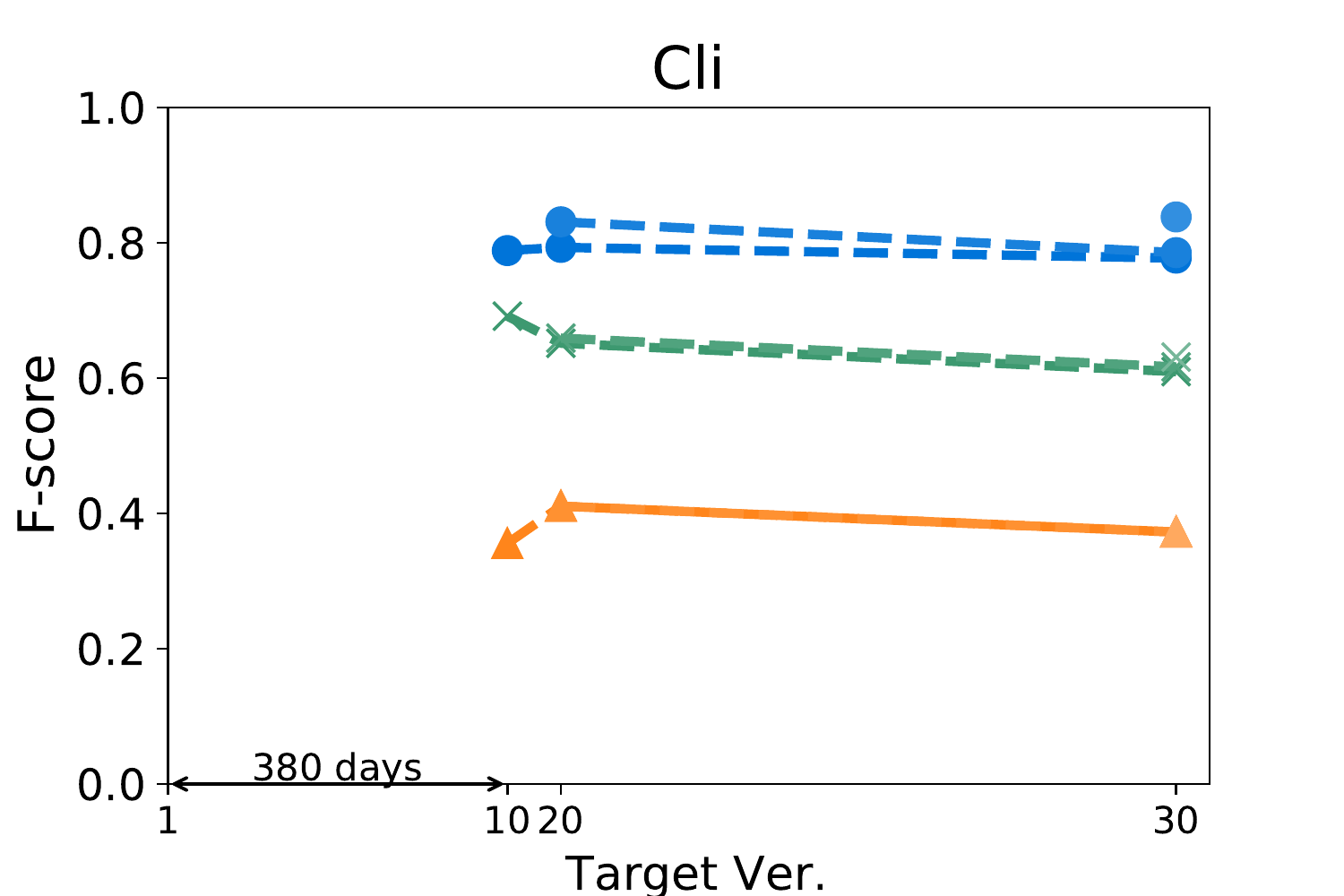}
    \includegraphics[width=0.42\textwidth]{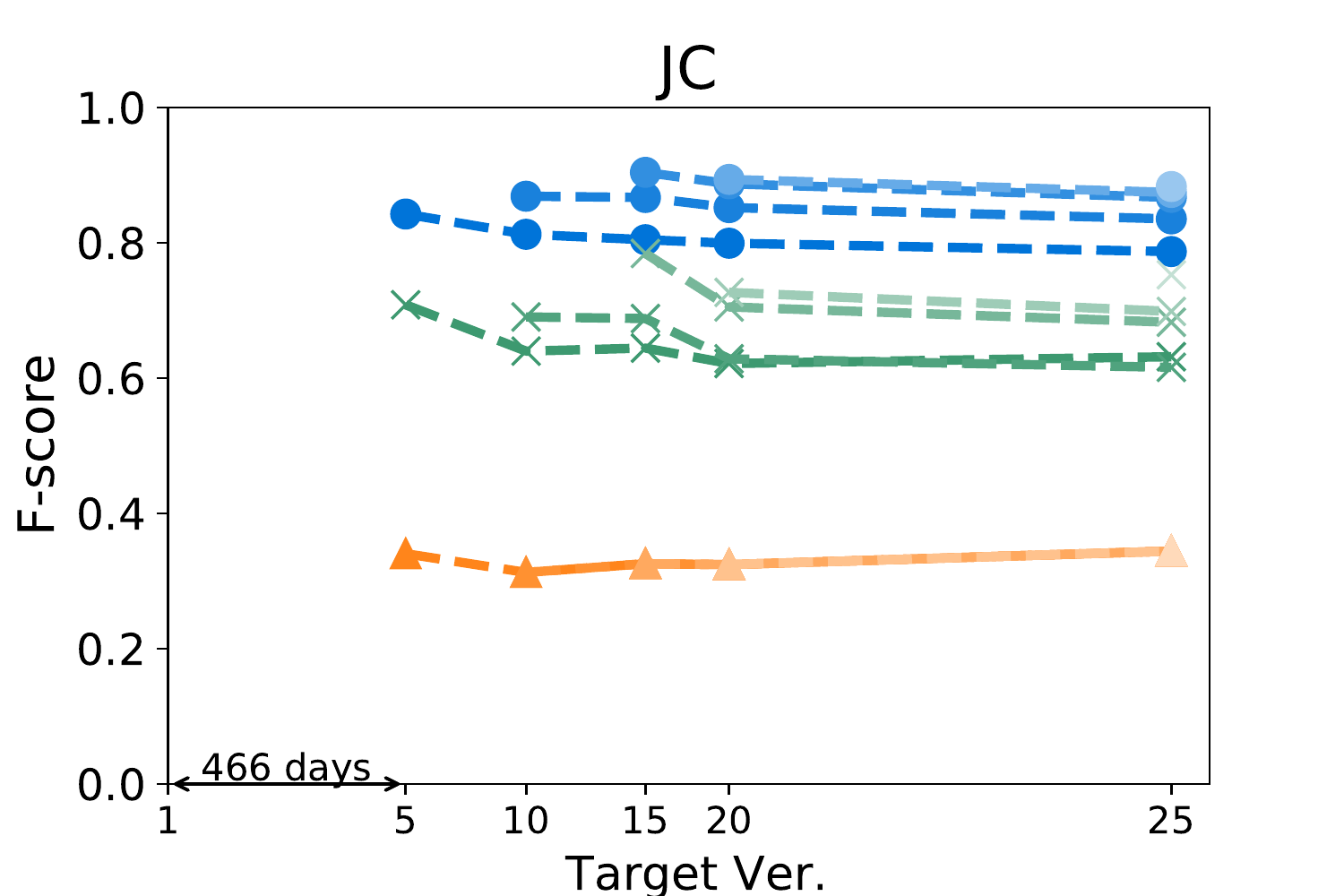}
    \includegraphics[width=0.42\textwidth]{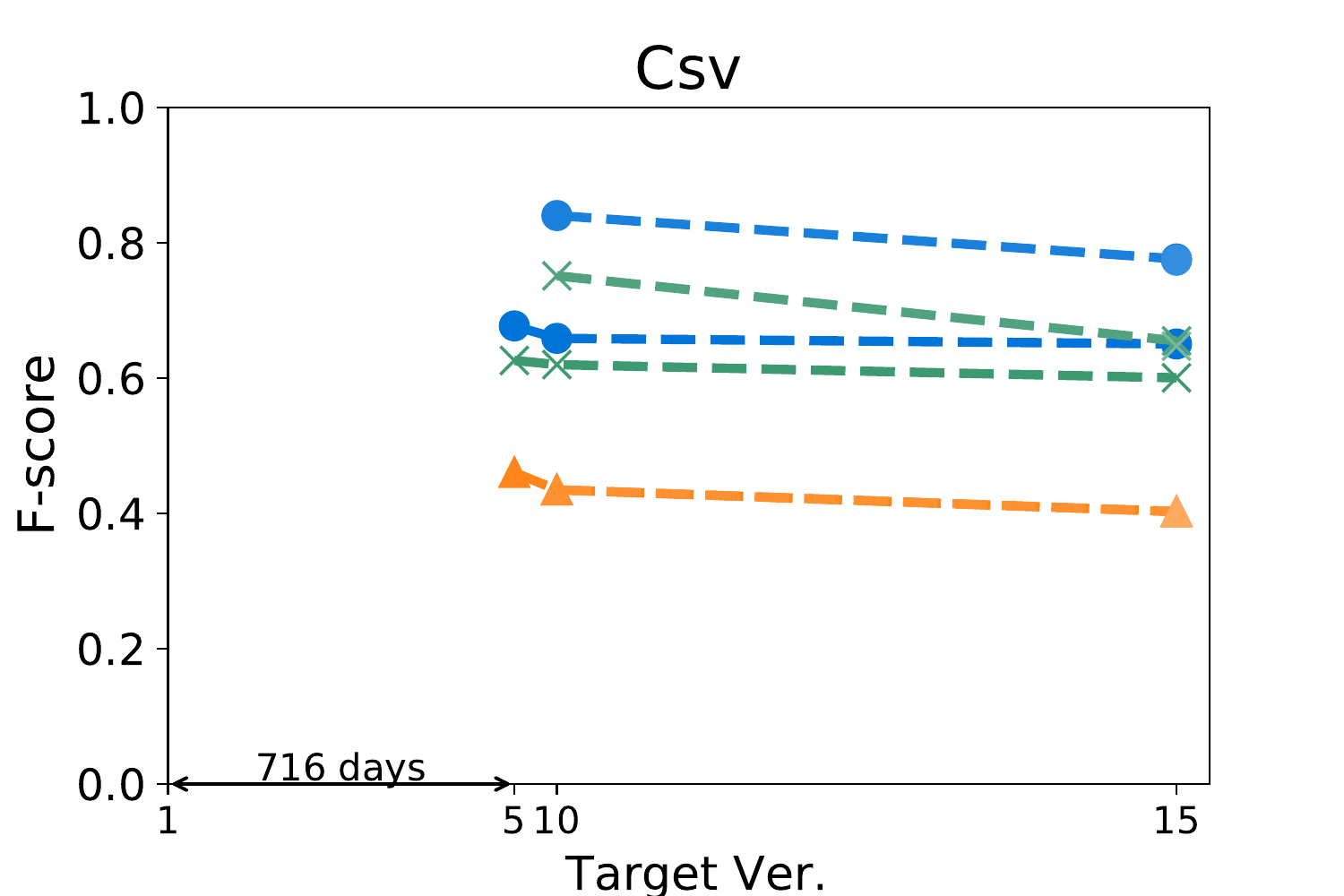}

    \caption{Prediction of the full kill matrix on PIT}
    \label{fig:RQ1_pit}
\end{figure}

\section{Results}
\label{sec:result}

This section answers the research questions using the results from the 
empirical evaluation of \name.

\subsection{Effectiveness (RQ1)}
\label{sec:rq1}

Figures~\ref{fig:RQ1_major} and \ref{fig:RQ1_pit} show how the F-score changes
when we train a model using a kill matrix of a base version and predict kill
matrices of subsequent versions that are gradually farther away. \add{The
$x$-axis shows the relative time interval between versions: the longer the time interval
between versions, the further the distance in $x$-axis between them. We also
specify the exact number of elapsed days between the first two versions in the
$x$-axis for the sake of understanding.} The colours in
Figures~\ref{fig:RQ1_major} and \new{\ref{fig:RQ1_pit} represent the models we
use: the blue circle marker represents \name, the green X marker represents PMT,
and the orange triangle marker represents the coverage based baseline. In the
same colour variation, the model trained with the older program is marked as
darker colour, and the connected line between markers represent that the
models are trained using the same base version.}

Figure~\ref{fig:RQ1_pit} shows that \name performs the best for versions of
Chart on PIT: the average F-score from all pairs of versions of Chart is 0.92.
For all subjects in PIT, \name achieves the average F-score 0.84. Compared to
PIT, as shown in Figure~\ref{fig:RQ1_major}, \name shows slightly worse
prediction performance on Major, with the overall average F-score 0.81.

\new{Also, \name outperforms PMT and the coverage based model. PMT and the coverage based
model produce average F-scores of 0.70 and 0.39, respectively, when predicting
the kill matrices of PIT, and 0.68 and 0.19 for the kill matrices of Major. This
supports our hypothesis that PMT lacks fine-grained features needed for test
case level prediction. It is not surprising that the coverage based heuristic
performs the worst, since the coverage is not a sufficient condition for killing
a mutant.}

By comparing lines with different tints in the same colour variation, we can
observe how the time between two versions affects the prediction accuracy of
\name. The older the model used for training is, the worse the prediction
becomes. When predicting for the latest version, models trained on the
immediately preceding studied version shows 0.08 and 0.09 point higher F-scores
when compared to models trained on the oldest versions, for Major and PIT
respectively. However, we also note that the accuracy degradation is relatively
slow: the F-scores decrease by 0.019 and 0.013 point per year on average for
Major and PIT, respectively. \new{PMT also follows this trend but the coverage based
model does not.}

The worst F-scores are observed in the models that are trained with the oldest 
versions of Gson and Csv. With both projects, the first interval is much 
longer than those in others: over five years for Gson, and almost two years 
for Csv. Combined with the relatively smaller size of these two projects, we 
suspect that the codebases for these two projects have changed more than 
other projects, which makes the prediction harder. For example, the oldest 
version of Gson only produces about half the mutants then those produced from 
the next analysed version, resulting in F-scores consistently lower than  0.7 
for all predictions based on itself.

\new{\textbf{Answer to RQ1:} \name successfully learns the features in the NL 
channel that are effective in predicting the full kill matrices: \name
achieves average F-score of 0.81 and 0.84 for Major and PIT, respectively, 
which significantly outperforms PMT and the coverage based model.}

\begin{table}[ht]
    \centering
    \caption{Efficiency of \name}
    \label{table:RQ2}
    \scalebox{0.9}{
        \begin{tabular}{ll|rrr||rrr}
            \toprule
            Project        & Ver.     & Major   & \name  & Speed-up   & PIT   & \name &
            Speed-up  \\
            \midrule
            Lang & 1 & 12,924s & \textbf{267s} & 48.34X & 1,472s & \textbf{128s} & 11.46X \\
            & 10 & 13,185s & \textbf{244s} & 54.10X & 1,385s & \textbf{116s} & 11.95X \\
            & 20 & 5,395s & \textbf{214s} & 25.15X & 970s & \textbf{104s} & 9.30X \\
            & 30 & 5,220s & \textbf{212s} & 24.61X & 900s & \textbf{97s} & 9.24X \\
            & 40 & 4,756s & \textbf{206s} & 23.08X & 836s & \textbf{93s} & 9.02X \\
            & 50 & 6,793s & \textbf{196s} & 34.72X & 865s & \textbf{93s} & 9.29X \\
            \midrule
            Time & 1 & - & - & - & 2,294s & \textbf{616s} & 3.73X \\
            & 5 & - & - & - & 2,170s & \textbf{592s} & 3.67X \\
            & 10 & - & - & - & 1,957s & \textbf{614s} & 3.19X \\
            & 15 & - & - & - & 2,138s & \textbf{564s} & 3.79X \\
            & 20 & - & - & - & 2,309s & \textbf{557s} & 4.15X \\
            \midrule
            Chart & 1 & 64,719s & \textbf{1,248s} & 51.87X & 2,295s & \textbf{369s} & 6.22X \\
            & 5 & 53,986s & \textbf{1,093s} & 49.40X & 2,014s & \textbf{338s} & 5.95X \\
            & 10 & 46,983s & \textbf{998s} & 47.07X & 1,542s & \textbf{293s} & 5.26X \\
            & 15 & 46,429s & \textbf{962s} & 48.27X & 1,520s & \textbf{289s} & 5.26X \\
            & 20 & 42,475s & \textbf{873s} & 48.68X & 1,466s & \textbf{258s} & 5.68X \\
            \midrule
            Gson & 15 & 16,738s & \textbf{347s} & 48.27X & 376s & \textbf{118s} & 3.19X \\
            & 10 & 15,986s & \textbf{339s} & 47.13X & 351s & \textbf{120s} & 2.94X \\
            & 5 & 15,253s & \textbf{348s} & 43.82X & 345s & \textbf{113s} & 3.04X \\
            \midrule
            Cli & 30 & 1,290s & \textbf{35s} & 36.98X & 74s & \textbf{19s} & 3.81X \\
            & 20 & 498s & \textbf{37s} & 13.35X & 43s & \textbf{20s} & 2.12X \\
            & 10 & 408s & \textbf{32s} & 12.94X & 42s & \textbf{18s} & 2.30X \\
            \midrule
            JC & 25 & 113,343s & \textbf{538s} & 210.53X & 1,391s & \textbf{204s} & 6.81X \\
            & 20 & 88,075s & \textbf{392s} & 224.44X & 1,013s & \textbf{142s} & 7.16X \\
            & 15 & 45,069s & \textbf{317s} & 142.17X & 669s & \textbf{115s} & 5.82X \\
            & 10 & 44,110s & \textbf{305s} & 144.85X & 676s & \textbf{113s} & 5.99X \\
            & 5 & 31,257s & \textbf{209s} & 149.49X & 557s & \textbf{83s} & 6.72X \\
            \midrule
            Csv & 15 & 5,289s & \textbf{52s} & 101.00X & 1,781s & \textbf{23s} & 77.91X \\
            & 10 & 1,317s & \textbf{33s} & 40.30X & 1,359s & \textbf{16s} & 83.12X \\
            & 5 & 1,179s & \textbf{31s} & 37.44X & 1,493s & \textbf{16s} & 93.90X \\
            \bottomrule
        \end{tabular}
    }
\end{table}

\subsection{Efficiency (RQ2)}
\label{sec:rq2}

The main drawback of building a full kill matrix comes from its huge
computational cost. It requires even more time than traditional mutation
testing, because even if a mutant is killed by some test earlier, it does not
skip running other tests. For example, considering Lang 10, traditional mutation
testing using PIT was 4.3x faster than those with full kill matrix option.
Therefore, we investigate how much execution time can be saved by \name,
compared to that of traditional mutation analysis with full kill matrix option.

Table~\ref{table:RQ2} shows the execution time of two mutation tools and \name
to compute the full kill matrix. Columns 3 and 4 list the execution time of Major
and \name that predicts Major's kill matrix of the corresponding version,
respectively. Similarly, Columns 6 and 7 list the execution time of PIT and \name.
Overall, the results show that PIT can compute the full kill matrix in a reasonable
time: it takes up to 38 minutes for Time and up to 74 seconds for Cli. Moreover,
PIT is on average 28x faster than Major since  
Major has generated more mutants than PIT in our study. Still, \name can speed up
from 2x to 94x, compared to PIT. In particular, \name predicts kill matrices of
Csv exceptionally faster than others; although all Csv versions have fewer than
300 tests, they are executed relatively slow, resulting in PIT running more
than 22 minutes. In contrast, \name is not affected by the test executions, but
only the number of tests, so it can predict kill matrices in less than 23
seconds. \name also requires significantly less time than Major for all 
subjects, with the average speed-up of 68x. Major runs more than 31 
hours for JC 25, but \name takes only nine minutes.

\add{Note that the reported speed-ups are specifically in the context of 
predicting the entire kill matrices. It is theoretically possible that \name 
can be even more efficient if we only wanted to predict the mutation scores like 
PMT: we simply need to stop the prediction for a mutant once it is predicted to 
be killed by any test case. However, as \name maps one-to-one relation between 
multiple mutants and test cases, the inference phase is fully parallelised 
using a GPU across the elements in the kill matrices, making mutation score 
level evaluation of efficiency difficult. Even if we serialise the computation, 
the ordering of prediction would affect when the ideal early stopping point 
should be. Consequently, we limit our evaluation of efficiency to the 
prediction of entire kill matrices.}

\textbf{Answer to RQ2:} \name is much more efficient in 
predicting the full kill matrix than traditional mutation analysis, with the 
average speed-up of 68x against Major and 14x against PIT.

\begin{figure}[!ht]
    \centering
    \begin{minipage}[t]{0.8\linewidth}
        \centering
        \includegraphics[width=\textwidth, trim=0 0.4cm 0 0, clip]{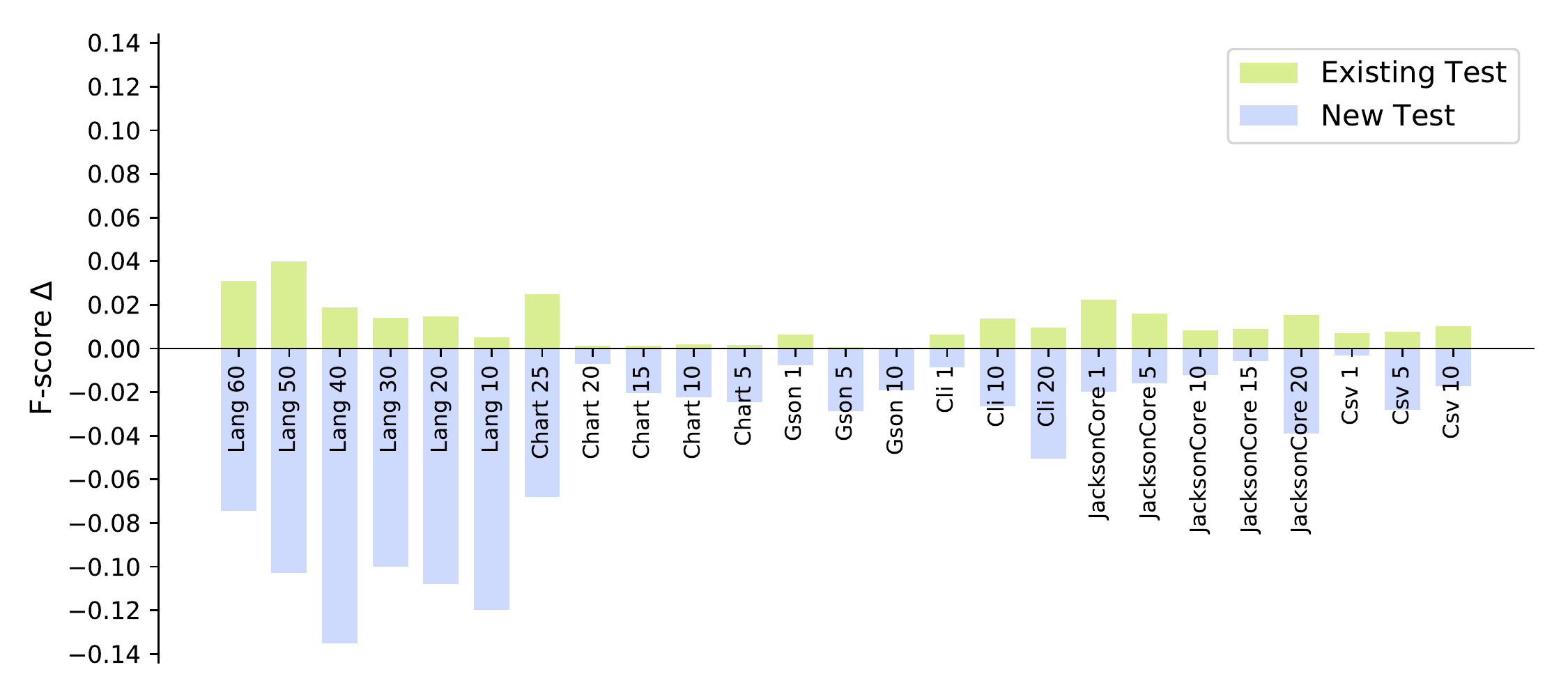}
        \subcaption{Major}
    \end{minipage}
 
    \begin{minipage}[t]{0.8 \linewidth}
        \centering
        \includegraphics[width=\textwidth, trim=0 0.4cm 0 0, clip]{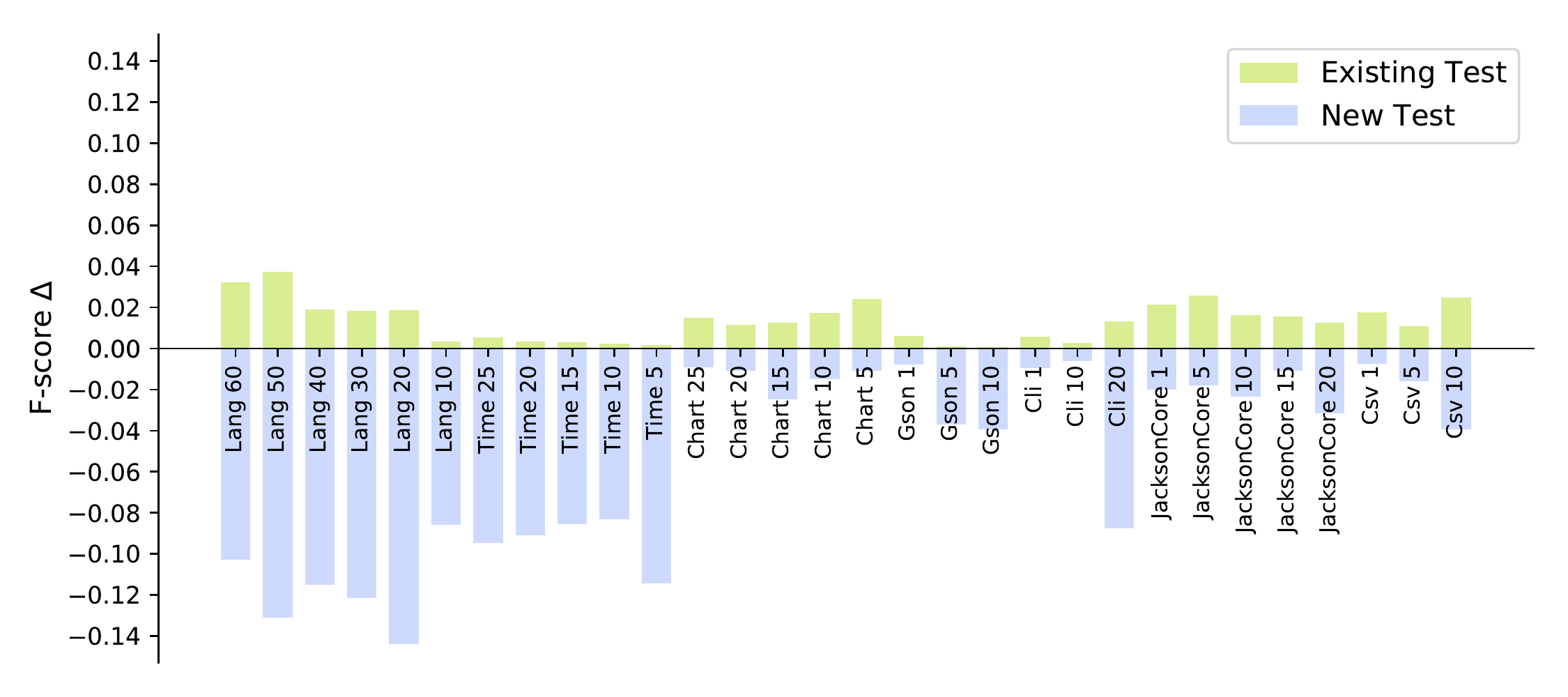}
        \subcaption{PIT}
    \end{minipage}
    \caption{Changes of F-score of existing tests and new tests}
    \label{fig:RQ3}
\end{figure}

\subsection{Generalisation (RQ3)}
\label{sec:rq3}

In RQ3, we investigate whether \name can generalise to the unseen test cases 
that did not exist in the test suite of the version used for training. Before 
prediction, we separate the test cases into two categories, existing and new. 
Subsequently, we report the prediction F-score for each group independently. 
While we only categorise test cases, we posit that the new test cases are 
likely to cover newly introduced code in the version that we apply \name to.

Figure~\ref{fig:RQ3} shows how F-score changes between the old and the new 
test categories per projects. The $x$-axis represents the version used 
for training, and the $y$-axis represents the averaged difference in the F-scores when 
both categories are compared to the F-score computed using all tests: the F-score 
$\Delta$ of 0 means that the category produces the same F-score as that of the 
entire test suite.

For all subjects, the existing tests show F-scores that are higher by 0.01 and 0.01
on average, while new tests show F-scores that are lower by 0.04  and 0.05 on
average, Major and PIT respectively. The difference suggests that \name does
perform better for the existing test cases. The notable outliers are from the versions
in Lang, for both Major and PIT: the average F-score $\Delta$ is
$-$0.11. We suspect that this is related to the
test case granularity: test cases in Lang tend to cover fewer methods 
compared to other projects. In our analysis, a test case in Lang covers about 14
methods on average, whereas a test case in all the other projects covers about
127 methods on average. The finer granularity may drive \name to learn
stronger one-to-one mappings between the test cases and the source methods, with higher
risk of overfitting. In turn, this will make it more difficult for \name to make
an accurate prediction for the new test cases. Another outlier is the versions in Time analysed with
PIT; however, the reason may be the opposite because Time has the highest average
number of methods covered by a test case, which is 360. It is more difficult to
learn mappings between the source and the test methods if the relationship is
dominantly one-to-many, resulting in low performance of \name.

\textbf{Answer to RQ3:} \name can generalise to the new source and test code: the 
F-score decreases only by 0.04 and 0.05 with Major and PIT, respectively, for 
new and unseen test cases.

\subsection{Mutation Score (RQ4)}
\label{sec:rq4}

\add{The test case level of PMA is finer-grained than the test suite level of
PMT: once we predict a kill matrix, we can easily compute the Mutation Score
(MS) from the matrix.} We expect a conversion in the opposite direction to be
more challenging. This is because, intuitively, PMA is more challenging as a
learning problem than PMT, since it needs to learn and produce more information
(an entire kill matrix) compared to PMT (which essentially predicts an
aggregation of a kill matrix).

\begin{table}[ht]
    \centering
    \caption{Mutation Score Prediction by \name and PMT}
    \label{table:pmt_ms}
    
    \begin{tabular}{l|rr|rr||rr|rr}
        \toprule
        \multirow{3}{*}{Project}  & \multicolumn{4}{c||}{Major} & \multicolumn{4}{c}{PIT} \\ \cline{2-9}
                    & \multicolumn{2}{c|}{F-Score} & \multicolumn{2}{c||}{MS Error} & \multicolumn{2}{c|}{F-Score} & \multicolumn{2}{c}{MS Error} \\ \cline{2-9}
                    & \name   & PMT   & \name & PMT & \name   & PMT   & \name & PMT \\           
        \midrule
        Lang & \textbf{0.825} & 0.715 & \textbf{0.848} & 17.547 & 0.908 & \textbf{0.912} & \textbf{2.112} & 4.503\\
        Time & - & - & - & - & 0.894 & \textbf{0.903} & \textbf{0.254} & 6.091\\
        Chart & \textbf{0.818} & 0.747 & \textbf{3.047} & 4.641 & \textbf{0.896} & 0.827 & \textbf{1.785} & 6.306\\
        Gson & 0.625 & \textbf{0.644} & 10.837 & \textbf{9.590} & \textbf{0.817} & 0.795 & \textbf{4.281} & 6.445\\
        Cli & 0.669 & \textbf{0.781} & 23.943 & \textbf{7.922} & 0.847 & \textbf{0.905} & 10.813 & \textbf{3.583}\\
        JC & \textbf{0.775} & 0.693 & \textbf{4.370} & 9.755 & \textbf{0.851} & 0.832 & \textbf{0.595} & 5.862\\
        Csv & \textbf{0.672} & 0.668 & 17.286 & \textbf{15.007} & 0.844 & \textbf{0.856} & \textbf{1.166} & 7.149\\
        \bottomrule
    \end{tabular}
    
\end{table}

Table~\ref{table:pmt_ms} shows the F-score and MS error (i.e., $|MS_{Real} -
MS_{Pred}|$) when \name and PMT predict whether a mutant is killed by the
entire test suite. In general, it is hard to determine which model performs
better because the differences of both F-score and MS error vary considerably
between the subjects. However, despite the aggregation over the predicted kill
matrix to compute MS, \name shows comparable and sometimes better results than
PMT.

In addition to the analysis of the absolute error, we further investigate the
actual difference (i.e., $MS_{Real} - MS_{Pred}$) for all predictions by \name
and PMT. In total, out of 55 predictions, 11 $MS_{Pred}$ by \name are higher 
than $MS_{Real}$. In contrast, all 55 $MS_{Pred}$ by PMT are higher than 
$MS_{Real}$. This indicates that PMT tends to overestimate mutation scores, 
while the errors of \name are both over and underestimations. For 
\name, the observed trend is plausible, because some of the kill relations may 
be very unique, and therefore harder to learn properly. In comparison, 
considering the importance of dynamic features for PMT, the overestimation by 
PMT is in line with the widely known limitations of structural coverage: it 
relies significantly on features related to the execution of the mutant, but 
the coverage does not necessarily result in killing it.

A hybrid model that combines the features of \name and PMT may improve the
prediction performance for the MS, as well as the contents of the
kill matrix. However, the focus of this paper was to evaluate the feasibility 
of \name based on the NL channel. We leave further investigation of 
hydridisation as future work.

\textbf{Answer to RQ4:} \name can predict the mutation results with average 
F-score of 0.80, and average mutation score error of 6.26, producing 
comparable results with PMT that achieves 0.79 and 8.03 respectively.

\subsection{Application Study (RQ5)}
\label{sec:RQ5}

\begin{table}[!ht]
    \centering
    \caption{A method-level fault localisation of four MBFL techniques}
    \label{table:RQ5}
    \scalebox{0.9}{
    \begin{tabular}{l|rr|rrrr}
        \toprule
        Technique & Project & \# Faults & $acc@1$ & $acc@3$ & $acc@5$ & $acc@10$ \\
        \midrule
        
        \multirow{5}{*}{MUSE~\cite{Moon:2014ly}} & Lang & 64 & 23 & 36 & 42 & 47 \\
        & Math & 105 & 20 & 42 & 51 & 64 \\
        & Time & 26 & 3 & 5 & 6 & 9 \\
        & Chart & 25 & 11 & 12 & 14 & 18 \\
         \cmidrule{2-7}
         & Total & 220 & 57 & 95 & 113 & 138 \\
        \midrule
        
        \multirow{5}{*}{Metallaxis~\cite{Papadakis:2015sf}} & Lang & 64 & 30 & 44 & 48 & 57 \\
        & Math & 105 & 22 & 49 & 60 & 74 \\
        & Time & 26 & 10 & 12 & 15 & 20 \\
        & Chart & 25 & 4 & 10 & 12 & 13 \\
         \cmidrule{2-7}
         & Total & 220 & 66 & 115 & 135 & 164 \\
         
        \midrule
        
        \multirow{5}{*}{SIMFL~\cite{kim2021issre}} & Lang & 64 & 47 & 54 & 55 & 55 \\
         & Math & 105 & 44 & 60 & 69 & 81 \\
         & Time & 26 & 11 & 16 & 18 & 23 \\
         & Chart & 25 & 9 & 13 & 15 & 19 \\
         \cmidrule{2-7}
         & Total & 220 & 111 & 143 & 157 & 178 \\
         
        \midrule
         
        \multirow{5}{*}{SIMFL with \name} & Lang & 64 & 36 & 48 & 52 & 56 \\
         & Math & 105 & 37 & 63 & 75 & 84 \\
         & Time & 26 & 8 & 11 & 12 & 13 \\
         & Chart & 25 & 7 & 14 & 19 & 22 \\
         \cmidrule{2-7}
         & Total & 220 & 88 & 136 & 158 & 175 \\
         
        \bottomrule
    \end{tabular}
    }
\end{table}

\add{In addition to evaluating the precision of \name, we design an application
study showing its usefulness when it is applied to other domains; we employ Fault
Localisation (FL) problem that aims to find a location of the faults. Our target
tool is SIMFL, Mutation Based Fault Localisation (MBFL) technique that
leverages the kill matrix to locate faults~\cite{kim2021issre}. By replacing the
kill matrix used in SIMFL with the predicted one by \name, we investigate how
much the localisation accuracy of SIMFL changes. Also, we present a comparison
to the two other MBFL techniques, MUSE and Metallaxis, using 220 buggy programs
in \dfj, as shown in Table~\ref{table:RQ5}. `SIMFL' refers to the SIMFL with its
own assumption utilising the original kill matrix, and `SIMFL with \name' refers
to SIMFL that uses the predicted kill matrix. In this study, we only used PIT
mutation tool as we failed to get the original kill matrix of Math
(commons-math) and Time using Major.\footnote{We have include commons-math of
\dfj for better comparison of the FL results to other techniques; running Major
on Time causes timeout after 48 hours.} \name models are trained on the oldest
version of the project, e.g., the model used to infer the kill matrices of Chart
1 to Chart 25 is trained on Chart 26. As FL techniques usually produce the ranks
of the suspicious program elements, the evaluation metric we use is $acc@n$, the
number of buggy programs where the MBFL technique successfully located its
faults within the top n position.}

\add{Table~\ref{table:RQ5} reports that SIMFL outperforms others: it locates 111
faults at the top, followed by SIMFL with \name, Metallaxis, and MUSE. Note that
SIMFL with \name still outperforms MUSE and Metallaxis for all $acc@n$ metrics
despite using predicted kill matrices. Moreover, compared to SIMFL, the
degradation of localisation effectiveness is relatively small for $acc@3$ and
$acc@10$ and it performs better with $acc@5$. It indicates that the predicted
kill matrices are sufficiently accurate for the task of FL, especially with
respect to $acc@10$ metric, although higher accuracy is needed to achieve
competitive $acc@5$ and above.}

\add{\textbf{Answer to RQ5:} the predicted kill matrix by \name is successfully
applied to the MBFL technique, SIMFL. SIMFL with \name outperforms other two
MBFL techniques and does not show disruptive degradation compared to the
original SIMFL assumption.}

\subsection{Naming Convention and Automated Test Generation (RQ6)}
\label{sec:RQ6}

\begin{table}[!ht]
    \centering
    \caption{Predicting kill matrix by replacing developer-written tests (Dev) with EvoSuite tests (Evo)}
    \label{table:RQ6}
    \begin{tabular}{ll|rrr|r}
        \toprule
        Project & Target & Dev$_{train}$ & Evo$_{train}$ & Evo$_{train}$ & Dev$_{train}$ \\
                & Vesrion & Evo$_{test}$ & Dev$_{test}$ & Evo$_{test}$ & Dev$_{test}$ \\
        \midrule
        Lang & 60 & 0.61 & 0.68 & 0.74 & 0.86 \\
        & 50 & 0.61 & 0.67 & 0.72 & 0.84  \\
        & 40 & 0.58 & 0.64 & 0.76 & 0.87  \\
        & 30 & 0.62 & 0.62 & 0.71 & 0.87  \\
        & 20 & 0.64 & 0.65 & 0.77 & 0.86  \\
        & 10 & 0.65 & 0.67 & 0.76 & 0.88  \\
        \midrule
        Time & 25 & 0.65 & 0.66 & 0.77 & 0.84 \\
        & 20 & 0.66 & 0.64 & 0.76 & 0.85  \\
        & 15 & 0.66 & 0.65 & 0.74 & 0.84  \\
        & 10 & 0.65 & 0.64 & 0.73 & 0.83  \\
        & 5 & 0.66 & 0.66 & 0.76 & 0.85  \\
        \midrule
        Gson & 1 & 0.56 & 0.60 & 0.67 & 0.66 \\
        & 5 & 0.72 & 0.64 & 0.82 & 0.90  \\
        & 10 & 0.71 & 0.66 & 0.81 & 0.91  \\
        \midrule
        Cli & 1 & 0.57 & 0.59 & 0.52 & 0.79 \\
        & 10 & 0.56 & 0.60 & 0.87 & 0.83  \\
        & 20 & 0.57 & 0.60 & 0.86 & 0.84  \\
        \midrule
        Csv & 1 & 0.57 & 0.58 & 0.66 & 0.68 \\
        & 5 & 0.68 & 0.71 & 0.76 & 0.84  \\
        & 10 & 0.57 & 0.68 & 0.70 & 0.77  \\
        \bottomrule
    \end{tabular}
\end{table}

\add{We evaluate \name with EvoSuite~\cite{Fraser:2013vn} generated test cases
in RQ6: the aim is to investigate whether the NL channel still holds useful
information when test cases are automatically generated. Instead of relying on
naming conventions of human developers, we use the descriptive naming strategy
of EvoSuite~\cite{Daka2017zf}. We hereafter denote developer-written test suites
with Dev, and EvoSuite generated test suites with Evo. We first collect kill
matrices of all Evo test suites. Subsequently, we train \name model on Evo
(i.e., Evo$_{train}$) and test on Dev (i.e., Dev$_{test}$), or train \name model
on Dev (i.e., Dev$_{train}$) and infer kill matrix for Evo (i.e., Evo$_{test}$).
In addition, we provide the results of Evo$_{train}$ \& Evo$_{test}$ and
Dev$_{train}$ \& Dev$_{test}$. Note that we only conduct the experiment using
PIT because running Major on EvoSuite tests times out after 48 hours.}

\add{Table~\ref{table:RQ6} presents the F-scores for each train-test pair.  
Overall, Dev$_{train}$ \& Dev$_{test}$ pair achieves the highest F-scores,
followed by Evo$_{train}$ \& Evo$_{test}$ pair. We posit that two factors 
contribute to the performance of \name with Dev test suites. First, human 
written tests may provide richer semantic information in the NL channel. 
Second, Dev is a regression test suite, whereas Evo test suites are obtained by 
independent and separate runs of EvoSuite for each version. Since Dev test 
suites are more stable (i.e., many test cases overlap between versions), the 
prediction of kill matrices may be easier for Dev test suites. However, the 
descriptive naming strategy of EvoSuite does provide some information in the NL 
channel, allowing \name to achieve F-score of up to 0.86 under the 
configuration of Evo$_{train}$ \& Evo$_{test}$.}

\add{
In contrast, replacing Dev with Evo results in deterioration of the effectiveness 
of \name. Dev$_{train}$ \& Evo$_{test}$ and Evo$_{train}$ \& Dev$_{test}$ show 
similar results, achieving average F-scores 0.62 and 0.64 respectively. The 
differences in naming convention between training and testing make the kill matrix prediction more difficult.}

\add{
\textbf{Answer to RQ6:} As long as a meaningful naming convention is applied during test generation, \name can also predict kill matrices for automatically generated test cases. 
}

\section{Discussion}
\label{sec:discussion}

This section raises a few issues that we have identified while performing the 
empirical evaluation. We believe that these can lead into interesting future 
work.

\subsection{Test Quality and Coding Convention}
\label{sec:convention}

Suppose we take any test suite for which \name can successfully perform
PMA, and remove all assertions from its test cases. This will render any
prediction useless. What \name does is to reconstruct the relationship between
test cases and mutants in the NL channel, based on the training data.
Consequently, it is vulnerable to such manipulation. In contrast, if a test case
without any assertion always kills a mutant using an implicit oracle (such as
crashes), \name will duly learn this relationship. In practice, we expect \name
to be used to reduce and amortise the cost of mutation analysis, and not to
completely replace it. If there is a continuity in the coding and naming
convention in both source and test code, \name is likely to maintain its
prediction accuracy. If there is a meaningful change in test quality, either
improvement or decrease, the prediction accuracy will degrade. We expect this
deviation to be picked up by the regular application of concrete mutation 
testing.

\subsection{Data Imbalance in PMA}
\label{sec:imbalance}

\new{
The balance between the number of mutant-test pairs that result in kills 
and non-kills cannot be known in advance, and likely not to be perfectly 1:1. 
As \name is essentially solving a classification problem, here we consider the 
implications of data imbalance.

The text-based nature of \name prevents the use of resampling approaches such as SMOTE~\cite{Chawla2002ve} or
ADASYN~\cite{He2008ti}. Therefore, we investigate the impact of class imbalance
on \name using random over- and undersampling (allowing duplicates). We over- and undersample the training data to make the class 
ratio 1:1. Subsequently, we train new models and evaluate them on the same 
subjects. Compared to the results of RQ1, with Major, the average 
F-score increases by 0.026 and 0.025 when we use over- and undersampling, 
respectively. With PIT, however, F-score decreases by 0.001 with oversampling, 
and increases by 0.010 with undersampling.

The different responses to over- and undersampling between Major and PIT can be
attributed to the status of their initial imbalance. The initial class ratio
between killed and not killed mutants generated by Major across all studied 
versions is 0.396 on average. With PIT, the initial imbalance ratio is 
much minor at 0.985 on average: there are actually more killed mutants 
than not killed ones in some projects. We suspect that this leads to over- and 
undersampling having more random effects on the results.

We note that the difference in F-score between initial and resampled
results are not significant. Data imbalance will have more significant 
consequences if it only appears in the training data. However, due to the 
continuity in development, projects with imbalanced kill matrices are likely to 
exhibit the same imbalance in the future, resulting in relatively minor 
consequences in our cross-version evaluation.

Overall, we conclude that addressing the class imbalance issue can improve the
performance of \name in general, but the implications of the imbalance can be
subtly different depending on the choice of the mutation tool, its 
configuration, and the contents of the kill matrix itself.}

\begin{table}[ht]
    \centering
    \caption{Results of Ablation Study on Major}
    \label{table:ablation}
    \scalebox{1.0}{
        \begin{tabular}{l|r|r|r|r}
            \toprule
            \multirow{2}{*}{Project}  & \multicolumn{4}{c}{Decrease in F-score} \\ \cline{2-5}
                     & Method Name   & Mutated Line   & Before\&After  & Mutation Op.  \\           
            \midrule
            Lang & 0.009 & -0.003 & 0.041 & -0.011 \\
            Chart & -0.056 & -0.057 & -0.030 & -0.072 \\
            Gson & -0.010 & 0.028 & 0.025 & -0.014 \\
            Cli & 0.018 & -0.003 & 0.078 & 0.014 \\
            JC & -0.006 & 0.016 & 0.018 & -0.015 \\
            Csv & 0.050 & -0.015 & 0.142 & 0.045 \\

            \bottomrule
        \end{tabular}
    }
\end{table}

\subsection{Ablation Study}
\label{sec:ablation_study}

\new{
To investigate which feature in \name contributes most to the prediction, 
we conduct an ablation study by removing each feature one by one and training 
the model. We then repeat the RQ1 study with Major to compare the F-scores 
from the ablated models to those from the original model. 
Table~\ref{table:ablation} reports the decrease of F-score for each removed 
feature. The column ``Method Name'' refers to the model that omits ``Source 
Method Name'', and the column ``Before \& After'' refers to the model that 
omits both ``Before'' and ``After'' in the model architecture, shown in 
Figure~\ref{fig:model_architecture}. The columns ``Mutated Line'' and 
``Mutation Operator'' refer to the models with corresponding input components 
removed, respectively.

Interestingly, the results show that the model performance varies across the
subjects but do not always deteriorate: for the versions of Chart, the ablated
models perform better than the original model. However, the average difference
of the F-score is 0.008, which may be too small to precisely assess the relative
feature importance when considering the stochastic nature of training DNN models
using hyperparameters. We presume that our features share some common
information and complement each other, allowing the ablated model to retain its
predictive power. We leave the design of more destructive study to measure the
relative feature importance as future work.}

\begin{table}[!t]
    \centering
    \caption{The changes between versions reported by cloc, only considering changed Java files.}
    \label{table:diff_subjects}    
    
    \begin{tabular}{llrrrr}
        \toprule            
        Project & Versions & \# Commits & Added & Modified & Removed \\
        \midrule
        Lang & 1 $\leftrightarrow$ 10 & 189 & 3,932 & 6,436 & 244 \\
        & 10 $\leftrightarrow$ 20 & 413 & 7,829 & 2,836 & 1,456 \\
        & 20 $\leftrightarrow$ 30 & 552 & 4,965 & 1,120 & 1,834 \\
        & 30 $\leftrightarrow$ 40 & 286 & 45,689 & 2 & 44,357 \\
        & 40 $\leftrightarrow$ 50 & 444 & 7,819 & 1,250 & 10,540 \\
        & 50 $\leftrightarrow$ 60 & 166 & 2,333 & 435 & 464 \\
       \midrule
       Gson & 15 $\leftrightarrow$ 10 & 89 & 637 & 222 & 74 \\
        & 10 $\leftrightarrow$ 5 & 79 & 377 & 21 & 137 \\
        & 5 $\leftrightarrow$ 1 & 823 & 14,474 & 950 & 8,220 \\
       \midrule
       Time & 1 $\leftrightarrow$ 5 & 16 & 583 & 873 & 21 \\
        & 5 $\leftrightarrow$ 10 & 62 & 1,306 & 229 & 548 \\
        & 10 $\leftrightarrow$ 15 & 68 & 623 & 81 & 11 \\
        & 15 $\leftrightarrow$ 20 & 31 & 396 & 71 & 11 \\
        & 20 $\leftrightarrow$ 25 & 65 & 855 & 136 & 99,260 \\
       \midrule
       Chart & 1 $\leftrightarrow$ 5 & 186 & 13,081 & 3,205 & 3,350 \\
        & 5 $\leftrightarrow$ 10 & 349 & 9,598 & 9,122 & 1,181 \\
        & 10 $\leftrightarrow$ 15 & 50 & 1,046 & 917 & 196 \\
        & 15 $\leftrightarrow$ 20 & 163 & 7,087 & 1,218 & 3,842 \\
        & 20 $\leftrightarrow$ 25 & 31 & 1,527 & 652 & 110 \\
       \midrule
       Cli & 30 $\leftrightarrow$ 20 & 252 & 5,678 & 0 & 4,566 \\
        & 20 $\leftrightarrow$ 10 & 51 & 941 & 597 & 1,222 \\
        & 10 $\leftrightarrow$ 1 & 61 & 520 & 72 & 141 \\
       \midrule
       Csv & 15 $\leftrightarrow$ 10 & 410 & 2,453 & 1,209 & 383 \\
        & 10 $\leftrightarrow$ 5 & 53 & 435 & 219 & 221 \\
        & 5 $\leftrightarrow$ 1 & 489 & 2,350 & 497 & 675 \\
       \midrule
       JC & 25 $\leftrightarrow$ 20 & 304 & 10,684 & 459 & 725 \\
        & 20 $\leftrightarrow$ 15 & 150 & 7,522 & 755 & 3,387 \\
        & 15 $\leftrightarrow$ 10 & 142 & 1,092 & 194 & 1,037 \\
        & 10 $\leftrightarrow$ 5 & 260 & 6,387 & 360 & 1,130 \\
        & 5 $\leftrightarrow$ 1 & 263 & 3,383 & 2,127 & 2,096 \\
       \bottomrule
    \end{tabular}
    
\end{table}

\subsection{Relation to Regression Mutation Testing}
\label{sec:comparison_with_rts}

We introduce and discuss the differences between \name and Regression Mutation 
Testing (ReMT)~\cite{zhang2012regression}, as ReMT has a similar goal as \name, 
namely to reduce the cost of mutation analysis in the context of evolving 
programs. When considering successive commits of evolving systems, ReMT reuses 
mutation testing results from the previous version and selects only the subset 
of tests that are affected by the latest change to rerun and update the 
kill matrix. If the underlying Regression Test Selection (RTS) technique is 
sound and complete, ReMT can output the correctly updated kill matrix with the 
minimum effort required.

ReMT is the most efficient when successively applied to each and every version:
if intervals become longer, changes will accumulate, making ReMT increasingly
inefficient (i.e., it will have to execute more and more tests and mutants).
This phenomenon is known to exist in the context of regression test case
selection~\cite{Kim:2005hb}. ReMT also involves other cost and assumptions. For
example, to precisely select mutants whose kill outcome may change in the new
version, some techniques adopt static analysis techniques that may incur
additional cost; to determine the mutant kill accurately, we also need to 
assume that the current version is correct. 

\name can be complementary to ReMT, as the results show that it can provide
reasonable predictions of kill matrices across much longer intervals, incurring
very little cost, without assuming anything about the correctness of the current
version. As an extreme example, suppose we want to update the kill matrix of
Lang 60 to get the kill matrix of Lang 1, two versions with 1,590 and 2,291
tests respectively (see Table~\ref{table:subject}). Even if we do not consider
any modified tests between two versions, at least 701 \emph{newly introduced
tests} (44\%) will have to be run. We do not think that this is a reasonable
scenario for ReMT. Table~\ref{table:diff_subjects} shows the magnitudes of
accumulated changes between versions, listing the number of commits, added,
modified, and removed lines between two adjacent versions. The number of commits
between adjacent versions, on average, is 216, which is a wider interval than
ReMT's setting that only considers the consecutive commits.

\section{Threats to Validity}
\label{sec:threats}

The major threats to internal validity lie in the implementation of \name as 
well as the correctness of actual mutation analysis that provides training 
data to \name. The models are implemented using widely used frameworks that 
withstood public scrutiny, such as Scipy and PyTorch. For training data, we 
depend on the publicly available mutation analysis script provided by \dfj.

The threats to external validity concern the choice of subjects, programming 
language, and the mutation tools. Due to the nature of our technique, the 
results are dependent on the coding conventions adopted by the studied 
projects. We tried to use the most recent version of \dfj with more diverse 
projects to avoid overfitting to a specific coding style. Since writing 
informative names is regarded as best practice, we expect the NL channel in 
source code written in other programming languages will provide similarly rich 
information. We adopt two mutation tools, Major and PIT, to reduce the threats 
related to the choice of a mutation tool.

The threats to construct validity include evaluation metrics we used to draw 
the conclusions. All evaluation metrics are standard measures for 
classification tasks: precision, recall, and F-score.

\section{Related Work}
\label{sec:related_work}

Existing approaches for mutation testing cost reduction include mutant
selection, mutant execution acceleration, and mutant score
prediction~\cite{pizzoleto2019systematic}. Offutt et al.~\cite{Offutt1996TOSEM}
suggested to use a small set of sufficient mutation operators instead of using a
vast number of mutation operators. Namin et al.~\cite{Namin2008ICSE} advanced
mutation operator selection by leveraging a multi-objective linear regression
which can learn an optimal mutation operator set according to given corpus of
mutation testing results. To reduce the runtime of mutant executions, Zhang et
al.~\cite{Zhang2013ICSE} applied the idea of test case prioritization to
improving the rate of mutant kills. AccMon~\cite{Wang2017ICSE} prevented
redundant mutant executions by monitoring internal states and cutting off a
mutant execution if it fails to induce different infection states.
\add{Regression Testing Selection (RTS)~\cite{chen2018speeding}
presented a practical Regression Mutation Testing~\cite{zhang2012regression}, by
selecting relevant tests that should be rerun to determine whether the new
mutants are killed. Despite having a similar regression assumption with ours, we
note that the RTS approach is specifically aimed at reducing the mutation
execution cost, while \name takes one step further to provide a predictive
model. Consequently, the RTS approach can be combined with \name when \name
needs to periodically update its reference kill matrix by (re)running test
cases. Recently introduced Predictive Mutation Testing (PMT)~\cite{Zhang2018gq}
opened up a new dimension in mutation testing by seeking to predict whether a
mutant will be killed or not by the entire test suite based on the structural
information on the target program and the coverage information on the given test
suite: consequently, it does not require any mutant execution. \name, whoever,
aims for a finer granularity in its prediction by attempting to infer the entire
kill matrices, instead of only predicting whether a mutant is killed by the test
suite or not.}

\add{The finer-grained, kill matrix level prediction may be utilised in any
application that depends on the one-to-one relationship between mutants and
individual test cases. Test generation techniques that use the mutants as a
guidance are one of such applications. For example, to create test cases that
can kill Subsuming Higher Order Mutants (SHOM)~\cite{Harman2011pi}, one needs to
start from the entire kill matrices of First Order Mutants, which \name can
approximate. DEMiner~\cite{Kim2018gd} proposed to improve a concolic testing
based on the information of which mutation affects which test execution: if
there exists an NL channel in test cases, \name can be used to approximately
capture the relationship at a lower cost. Some existing Automated Program Repair
(APR) techniques have taken advantage the mutation analysis. Weimer et
al.~\cite{Weimer2013ma} highlighted Generate-and-Validate program repair as a
dual of the mutation testing, while PraPR~\cite{ghanbari2019practical}
explicitly exploited mutation tools for APR. The high cost of patch validation
in APR suggests that, in both cases, \name may partially replace concrete
mutation analysis to find an attractive trade-off between analysis cost and
accuracy.} 

\add{A downstream task that can directly benefit from PMA in particular is Mutation Based Fault Localisation
(MBFL)~\cite{Wong:2016aa}. MBFL techniques have exploited a mutation analysis
that relates source code and tests through mutants~\cite{kim2021issre,
Papadakis:2015sf, Moon:2014ly, Hong:2015db}. By introducing the syntactic
modifications (i.e., mutants) to the buggy program, MBFL techniques observe the
changes of test results then assign a suspiciousness score to each statement to
identify the possible location of the given faults. For instance,
MUSE~\cite{Moon:2014ly} and MUSEUM~\cite{Hong:2015db} are based on the two
conjectures: mutating correct statements is likely to make passed tests fail,
and mutating faulty statements is likely to make failed tests pass. Then they
compute the suspiciousness scores using the ratio of fail-to-pass tests and
pass-to-fail tests. Using Spectrum based Fault Localisation (SBFL)-like
formulas, Metallaxis~\cite{Papadakis:2015sf} mutates the buggy program and see
whether the test results show a similar pattern with the patterns of the faults.
SIMFL utilises a statistical inference on the kill matrix to find the mutant
that makes tests fail in a similar pattern with the faults, then it suspects
that the faults would be close to the location of the 
mutant~\cite{kim2021issre}. In all cases, the cost of mutation analysis is 
directly added to the cost of debugging, as these techniques tend to perform 
mutation analysis once a bug is detected. \name can be easily applied to them 
to reduce the cost of the mutation analysis.}

\add{Even though the mutation testing has been considered a robust measurement
of test effectiveness, code coverage has been widely used as an alternative 
proxy due to its efficiency. It measures how well the specific code structures 
are exercised by the tests, e.g., a statement coverage checks whether the line 
is executed by at least one test case. Several studies have shown that there is 
a correlation between code coverage and test effectiveness~\cite{
gligoric2013comparing,gopinath2014code}, which supports the use of the coverage 
as one of our baselines. However, as it does not convey the intent of the given 
test suite, further studies have found that the test size and the assertion 
should be taken into account to quantify test effectiveness~\cite{
inozemtseva2014coverage, zhang2015assertions}. Checked coverage~\cite{
schuler2013checked} is one such example that attempted to incorporate the 
information of the assertions in the coverage using backwards slicing on the 
assertions in the test. However, its dependence on the static analysis may limit 
the applicability of checked coverage.}

Moreover, our work is also different from the 
existing cost reduction techniques in that \name utilises the natural language 
channel between the target program and the test cases to construct the 
prediction model.
Recent improvements of neural language models have enabled various techniques 
to exploit the semantic information inferred from the natural language channel 
in programs~\cite{Hindle:2012kq,Ray2016lz,Casalnuovo2020oj}. Since the natural 
language channel is both related to and constrained by the algorithm 
channel~\cite{Casalnuovo2020oj}, a hybrid approach would be promising to 
overcome the limitation of any existing analysis that aims to reason about 
program semantics. For example, neural word embedding has been used to 
discover semantically similar code snippets~\cite{Gu2018ICSE}, as well as 
for translating API descriptions into corresponding formal 
specifications~\cite{Blasi2018ICSE}. Recurrent Neural Network (RNN) models 
have been used to predict type signatures of JavaScript functions based on 
function and parameter names and comments~\cite{Malik2019ICSE}. Similarly, 
\name aims to approximate the relationship between mutants and test cases that 
kill them based on the similarity between their names in the embedding space.

\section{Conclusion}
\label{sec:conclusion}

In this paper, we propose \name, a Predictive Mutation Analysis (PMA) technique
that can learn and predict \emph{an entire kill matrix}, as opposed to
Predictive Mutation Testing (PMT) whose aim is to predict the mutation score.
\name exploits a Natural Language channel in the source code and test, as well
as the mutant-specific code fragments to organise input features through Deep
Neural Network (DNN). Empirical evaluation on 37 subjects in \dfj and two widely
used Java mutation tools, Major and PIT, demonstrates that \name predicts kill
matrices with average F-score of 0.83 and is 39 times faster than traditional
mutation analysis. In addition to predicting existing tests precisely, \name
generalises to the new tests, with degradation of the prediction accuracy by
0.05 in F-score. \name outperforms PMT and a coverage based baseline model in
predicting of the entire kill matrix by 0.14 and 0.45 point in F-score, and
performs as well as PMT in predicting of the mutation score only. \add{An
application study on Mutation Based Fault Localisation (MBFL) technique shows
that \name is successfully applied to the MBFL technique called SIMFL and it
achieves competitive accuracy even when it uses the predicted kill matrices
instead of real ones. We also show that \name can exploit NL channel of automatically
generated test suite as long as a meaningful naming convention is upheld.}

\bibliographystyle{plain}
\bibliography{references}

\begin{thebibliography}{10}

\bibitem{Blasi2018ICSE}
Arianna Blasi, Alberto Goffi, Konstantin Kuznetsov, Alessandra Gorla,
  Michael~D. Ernst, Mauro Pezz\`{e}, and Sergio~Delgado Castellanos.
\newblock Translating code comments to procedure specifications.
\newblock In {\em Proceedings of the 27th ACM SIGSOFT International Symposium
  on Software Testing and Analysis}, ISSTA 2018, page 242–253, New York, NY,
  USA, 2018. Association for Computing Machinery.

\bibitem{Casalnuovo2020oj}
Casey Casalnuovo, Earl~T. Barr, Santanu~Kumar Dash, Prem Devanbu, and Emily
  Morgan.
\newblock A theory of dual channel constraints.
\newblock In {\em Proceedings of the 42nd IEEE/ACM International Conference on
  Software Engineering: New Ideas and Emerging Results}, ICSE NIER 2020, pages
  25--28, 2020.

\bibitem{Chawla2002ve}
Nitesh~V. Chawla, Kevin~W. Bowyer, Lawrence~O. Hall, and W.~Philip Kegelmeyer.
\newblock Smote: Synthetic minority over-sampling technique.
\newblock {\em Journal of Artificial Intelligence Research}, 16(1):321--357,
  June 2002.

\bibitem{chen2018speeding}
Lingchao Chen and Lingming Zhang.
\newblock Speeding up mutation testing via regression test selection: An
  extensive study.
\newblock In {\em 2018 IEEE 11th International Conference on Software Testing,
  Verification and Validation (ICST)}, pages 58--69. IEEE, 2018.

\bibitem{cho2014learning}
Kyunghyun Cho, Bart Van~Merri{\"e}nboer, Caglar Gulcehre, Dzmitry Bahdanau,
  Fethi Bougares, Holger Schwenk, and Yoshua Bengio.
\newblock Learning phrase representations using rnn encoder-decoder for
  statistical machine translation.
\newblock {\em arXiv preprint arXiv:1406.1078}, 2014.

\bibitem{Coles2016ft}
Henry Coles, Thomas Laurent, Christopher Henard, Mike Papadakis, and Anthony
  Ventresque.
\newblock Pit: A practical mutation testing tool for java (demo).
\newblock In {\em Proceedings of the 25th International Symposium on Software
  Testing and Analysis}, ISSTA 2016, pages 449--452, New York, NY, USA, 2016.
  Association for Computing Machinery.

\bibitem{Daka2017zf}
Ermira Daka, Jos\'{e}~Miguel Rojas, and Gordon Fraser.
\newblock Generating unit tests with descriptive names or: Would you name your
  children thing1 and thing2?
\newblock In {\em Proceedings of the 26th ACM SIGSOFT International Symposium
  on Software Testing and Analysis}, ISSTA 2017, pages 57--67, New York, NY,
  USA, 2017. Association for Computing Machinery.

\bibitem{Debroy2010oa}
V.~{Debroy} and W.~E. {Wong}.
\newblock Using mutation to automatically suggest fixes for faulty programs.
\newblock In {\em 2010 Third International Conference on Software Testing,
  Verification and Validation}, pages 65--74, 2010.

\bibitem{Debroy2014nf}
Vidroha Debroy and W.~Eric Wong.
\newblock Combining mutation and fault localization for automated program
  debugging.
\newblock {\em Journal of Systems and Software}, 90:45 -- 60, 2014.

\bibitem{Fraser:2013vn}
Gordon Fraser and Andrea Arcuri.
\newblock Whole test suite generation.
\newblock {\em IEEE Trans. Softw. Eng.}, 39(2):276--291, February 2013.

\bibitem{ghanbari2019practical}
Ali Ghanbari, Samuel Benton, and Lingming Zhang.
\newblock Practical program repair via bytecode mutation.
\newblock In {\em Proceedings of the 28th ACM SIGSOFT International Symposium
  on Software Testing and Analysis}, pages 19--30, 2019.

\bibitem{gligoric2013comparing}
Milos Gligoric, Alex Groce, Chaoqiang Zhang, Rohan Sharma, Mohammad~Amin
  Alipour, and Darko Marinov.
\newblock Comparing non-adequate test suites using coverage criteria.
\newblock In {\em Proceedings of the 2013 International Symposium on Software
  Testing and Analysis}, pages 302--313, 2013.

\bibitem{Gopinath2015nu}
R.~{Gopinath}, A.~{Alipour}, I.~{Ahmed}, C.~{Jensen}, and A.~{Groce}.
\newblock How hard does mutation analysis have to be, anyway?
\newblock In {\em 2015 IEEE 26th International Symposium on Software
  Reliability Engineering (ISSRE)}, pages 216--227, 2015.

\bibitem{gopinath2014code}
Rahul Gopinath, Carlos Jensen, and Alex Groce.
\newblock Code coverage for suite evaluation by developers.
\newblock In {\em Proceedings of the 36th International Conference on Software
  Engineering}, pages 72--82, 2014.

\bibitem{Gu2018ICSE}
Xiaodong Gu, Hongyu Zhang, and Sunghun Kim.
\newblock Deep code search.
\newblock In {\em Proceedings of the 40th International Conference on Software
  Engineering}, ICSE '18, page 933–944, New York, NY, USA, 2018. Association
  for Computing Machinery.

\bibitem{Harman2011pi}
Mark Harman, Yue Jia, and William~B. Langdon.
\newblock Strong higher order mutation-based test data generation.
\newblock In {\em Proceedings of the 19th ACM SIGSOFT Symposium and the 13th
  European Conference on Foundations of Software Engineering}, ESEC/FSE '11,
  pages 212--222, New York, NY, USA, 2011. Association for Computing Machinery.

\bibitem{He2008ti}
Haibo He, Yang Bai, E.~A. {Garcia}, and Shutao Li.
\newblock Adasyn: Adaptive synthetic sampling approach for imbalanced learning.
\newblock In {\em 2008 IEEE International Joint Conference on Neural Networks
  (IEEE World Congress on Computational Intelligence)}, pages 1322--1328, June
  2008.

\bibitem{hellendoorn2017deep}
Vincent~J Hellendoorn and Premkumar Devanbu.
\newblock Are deep neural networks the best choice for modeling source code?
\newblock In {\em Proceedings of the 2017 11th Joint Meeting on Foundations of
  Software Engineering}, pages 763--773, 2017.

\bibitem{Hindle:2012kq}
Abram Hindle, Earl~T. Barr, Zhendong Su, Mark Gabel, and Premkumar Devanbu.
\newblock On the naturalness of software.
\newblock In {\em Proceedings of the 34th International Conference on Software
  Engineering}, ICSE '12, pages 837--847, Piscataway, NJ, USA, 2012. IEEE
  Press.

\bibitem{hoang2020cc2vec}
Thong Hoang, Hong~Jin Kang, Julia Lawall, and David Lo.
\newblock Cc2vec: Distributed representations of code changes.
\newblock In {\em Proceedings of the 42nd International Conference on Software
  Engineering}, ICSE ’20. ACM, 2020.

\bibitem{Hong:2015db}
Shin Hong, Byeongcheol Lee, Taehoon Kwak, Yiru Jeon, Bongsuk Ko, Yunho Kim, and
  Moonzoo Kim.
\newblock Mutation-based fault localization for real-world multilingual
  programs {(T)}.
\newblock In {\em 30th {IEEE/ACM} International Conference on Automated
  Software Engineering, {ASE} 2015, Lincoln, NE, USA, November 9-13, 2015},
  pages 464--475, 2015.

\bibitem{howden82}
William~E. Howden.
\newblock Weak mutation testing and completeness of test sets.
\newblock {\em IEEE Transactions on Software Engineering}, 8:371--379, 1982.

\bibitem{hucka2018spiral}
Michael Hucka.
\newblock Spiral: splitters for identifiers in source code files.
\newblock {\em Journal of Open Source Software}, 3(24):653, 2018.

\bibitem{inozemtseva2014coverage}
Laura Inozemtseva and Reid Holmes.
\newblock Coverage is not strongly correlated with test suite effectiveness.
\newblock In {\em Proceedings of the 36th international conference on software
  engineering}, pages 435--445, 2014.

\bibitem{just2014major}
Ren{\'e} Just.
\newblock The major mutation framework: Efficient and scalable mutation
  analysis for java.
\newblock In {\em Proceedings of the 2014 International Symposium on Software
  Testing and Analysis}, pages 433--436, 2014.

\bibitem{Karampatsis2020ICSE}
Rafael~Michael Karampatsis, Hlib Babii, Romain Robbes, Charles Sutton, and
  Andrea Janes.
\newblock {Big Code != Big Vocabulary: Open-Vocabulary Models for Source code}.
\newblock In {\em Proceedings of the 42nd International Conference on Software
  Engineering}, ICSE ’20. ACM, 2020.

\bibitem{kim2021issre}
Jinhan Kim, Gabin An, Robert Feldt, and Shin Yoo.
\newblock Ahead of time mutation based fault localisation using statistical
  inference.
\newblock In {\em 32nd {IEEE} International Symposium on Software Reliability
  Engineering, {ISSRE} 2021}, 2021.

\bibitem{Kim:2005hb}
Jung-Min Kim, Adam Porter, and Gregg Rothermel.
\newblock An empirical study of regression test application frequency.
\newblock {\em Software {T}esting, {V}erification, and {R}eliability},
  15(4):257--279, 2005.

\bibitem{Kim2018gd}
Y.~Kim, S.~Hong, B.~Ko, D.~L. Phan, and M.~Kim.
\newblock Invasive software testing: Mutating target programs to diversify test
  exploration for high test coverage.
\newblock In {\em 2018 IEEE 11th International Conference on Software Testing,
  Verification and Validation (ICST)}, pages 239--249, April 2018.

\bibitem{Knuth1984ci}
Donald~E. Knuth.
\newblock Literate programming.
\newblock {\em Comput. J.}, 27(2):97--111, May 1984.

\bibitem{Malik2019ICSE}
Rabee~Sohail Malik, Jibesh Patra, and Michael Pradel.
\newblock Nl2type: Inferring javascript function types from natural language
  information.
\newblock In {\em Proceedings of the 41st International Conference on Software
  Engineering}, ICSE '19, page 304–315. IEEE Press, 2019.

\bibitem{Mao2019ur}
Dongyu Mao, Lingchao Chen, and Lingming Zhang.
\newblock An extensive study on cross-project predictive mutation testing.
\newblock In {\em 2019 12th IEEE Conference on Software Testing, Validation and
  Verification (ICST)}, pages 160--171. IEEE, 2019.

\bibitem{Moon:2014ly}
Seokhyeon Moon, Yunho Kim, Moonzoo Kim, and Shin Yoo.
\newblock Ask the mutants: Mutating faulty programs for fault localization.
\newblock In {\em Proceedings of the 7th International Conference on Software
  Testing, Verification and Validation}, ICST 2014, pages 153--162, 2014.

\bibitem{Offutt1996TOSEM}
A.~Jefferson Offutt, Ammei Lee, Gregg Rothermel, Roland~H. Untch, and Christian
  Zapf.
\newblock An experimental determination of sufficient mutant operators.
\newblock {\em ACM Trans. Softw. Eng. Methodol.}, 5(2):99–118, April 1996.

\bibitem{off_lee94}
A.~Jefferson Offutt and Stephen~D. Lee.
\newblock An empirical evaluation of weak mutation.
\newblock {\em IEEE Transactions on Software Engineering}, 20:337--344, 1994.

\bibitem{Papadakis2010sf}
M.~{Papadakis} and N.~{Malevris}.
\newblock Automatic mutation test case generation via dynamic symbolic
  execution.
\newblock In {\em 2010 IEEE 21st International Symposium on Software
  Reliability Engineering}, pages 121--130, 2010.

\bibitem{Papadakis2019aa}
Mike Papadakis, Marinos Kintis, Jie Zhang, Yue Jia, Yves~Le Traon, and Mark
  Harman.
\newblock Chapter six - mutation testing advances: An analysis and survey.
\newblock volume 112 of {\em Advances in Computers}, pages 275 -- 378.
  Elsevier, 2019.

\bibitem{Papadakis:2015sf}
Mike Papadakis and Yves~Le Traon.
\newblock Metallaxis-fl: mutation-based fault localization.
\newblock {\em Softw. Test., Verif. Reliab.}, 25(5-7):605--628, 2015.

\bibitem{pizzoleto2019systematic}
Alessandro~Viola Pizzoleto, Fabiano~Cutigi Ferrari, Jeff Offutt, Leo Fernandes,
  and M{\'a}rcio Ribeiro.
\newblock A systematic literature review of techniques and metrics to reduce
  the cost of mutation testing.
\newblock {\em Journal of Systems and Software}, 157:110388, 2019.

\bibitem{Ray2016lz}
Baishakhi Ray, Vincent Hellendoorn, Saheel Godhane, Zhaopeng Tu, Alberto
  Bacchelli, and Premkumar Devanbu.
\newblock On the "naturalness" of buggy code.
\newblock In {\em Proceedings of the 38th International Conference on Software
  Engineering}, ICSE '16, pages 428--439, New York, NY, USA, 2016. ACM.

\bibitem{schuler2013checked}
David Schuler and Andreas Zeller.
\newblock Checked coverage: an indicator for oracle quality.
\newblock {\em Software testing, verification and reliability}, 23(7):531--551,
  2013.

\bibitem{Namin2008ICSE}
Akbar Siami~Namin, James~H. Andrews, and Duncan~J. Murdoch.
\newblock Sufficient mutation operators for measuring test effectiveness.
\newblock In {\em Proceedings of the 30th International Conference on Software
  Engineering}, ICSE '08, page 351–360, New York, NY, USA, 2008. Association
  for Computing Machinery.

\bibitem{ToTTestname}
Andrew Trenk.
\newblock Testing on the toilet: Writing descriptive test names.
\newblock
  \url{https://testing.googleblog.com/2014/10/testing-on-toilet-writing-descriptive.html}.
\newblock Accessed: 2020-08-28.

\bibitem{untch:mutation}
Roland~H. Untch, A.~Jefferson Offutt, and Mary~Jean Harrold.
\newblock Mutation analysis using mutant schemata.
\newblock In Thomas Ostrand and Elaine Weyuker, editors, {\em Proceedings of
  the 1993 International Symposium on Software Testing and Analysis ({ISSTA})},
  pages 139--148, 1993.

\bibitem{Voas:1992uq}
Jeffrey~M. Voas.
\newblock {\em {IEEE Transactions on Software Engineering}}.

\bibitem{Wang2017ICSE}
Bo~Wang, Yingfei Xiong, Yangqingwei Shi, Lu~Zhang, and Dan Hao.
\newblock Faster mutation analysis via equivalence modulo states.
\newblock In {\em Proceedings of the 26th ACM SIGSOFT International Symposium
  on Software Testing and Analysis}, ISSTA 2017, page 295–306, New York, NY,
  USA, 2017. Association for Computing Machinery.

\bibitem{wang2016compare}
Shuohang Wang and Jing Jiang.
\newblock A compare-aggregate model for matching text sequences.
\newblock {\em arXiv preprint arXiv:1611.01747}, 2016.

\bibitem{Weimer2013ma}
W.~{Weimer}, Z.~P. {Fry}, and S.~{Forrest}.
\newblock Leveraging program equivalence for adaptive program repair: Models
  and first results.
\newblock In {\em 2013 28th IEEE/ACM International Conference on Automated
  Software Engineering (ASE)}, pages 356--366, 2013.

\bibitem{Wong:2016aa}
W.~E. Wong, Ruizhi Gao, Yihao Li, Rui Abreu, and Franz Wotawa.
\newblock A survey on software fault localization.
\newblock {\em IEEE Transactions on Software Engineering}, 42(8):707, August
  2016.

\bibitem{Yoo:2010fk}
Shin Yoo and Mark Harman.
\newblock Regression testing minimisation, selection and prioritisation: A
  survey.
\newblock {\em Software {T}esting, {V}erification, and {R}eliability},
  22(2):67--120, March 2012.

\bibitem{Zhang2018gq}
Jie Zhang, Lingming Zhang, Mark Harman, Dan Hao, Yue Jia, and Lu~Zhang.
\newblock Predictive mutation testing.
\newblock {\em IEEE Transactions on Software Engineering}, 45(9):898--918,
  2018.

\bibitem{Zhang2013ICSE}
Lingming Zhang, Darko Marinov, and Sarfraz Khurshid.
\newblock Faster mutation testing inspired by test prioritization and
  reduction.
\newblock In {\em Proceedings of the 2013 International Symposium on Software
  Testing and Analysis}, ISSTA 2013, page 235–245, New York, NY, USA, 2013.
  Association for Computing Machinery.

\bibitem{zhang2012regression}
Lingming Zhang, Darko Marinov, Lu~Zhang, and Sarfraz Khurshid.
\newblock Regression mutation testing.
\newblock In {\em Proceedings of the 2012 International Symposium on Software
  Testing and Analysis}, pages 331--341, 2012.

\bibitem{zhang2015assertions}
Yucheng Zhang and Ali Mesbah.
\newblock Assertions are strongly correlated with test suite effectiveness.
\newblock In {\em Proceedings of the 2015 10th Joint Meeting on Foundations of
  Software Engineering}, pages 214--224, 2015.

\end{thebibliography}
\end{document}